\documentclass[aps,preprint,nobibnotes,nofootinbib]{revtex4} 
\usepackage{amsfonts}
\usepackage{graphicx}
\usepackage{amsmath}
\usepackage{amssymb}
\usepackage{subfigure}
\usepackage{setspace}

\begin{document}

\title{A Rigorous Derivation of Gravitational Self-force}

\author{Samuel E. Gralla}
\author{Robert M. Wald}
\affiliation{\it Enrico
Fermi Institute and Department of Physics \\ \it University of Chicago
\\ \it 5640 S.~Ellis Avenue, Chicago, IL~60637, USA}

\begin{abstract}
There is general agreement that the MiSaTaQuWa equations should
describe the motion of a ``small body" in general relativity,
taking into account the leading order self-force effects. However,
previous derivations of these equations have made a number of ad hoc
assumptions and/or contain a number of unsatisfactory features. For
example, all previous derivations have invoked, without proper
justification, the step of ``Lorenz gauge relaxation'', wherein the
linearized Einstein equation is written down in the form appropriate
to the Lorenz gauge, but the Lorenz gauge condition is then not
imposed---thereby making the resulting equations for the metric
perturbation inequivalent to the linearized Einstein equations. (Such
a ``relaxation'' of the linearized Einstein equations is essential in
order to avoid the conclusion that ``point particles'' move on
geodesics.) In this paper, we analyze the issue of ``particle motion''
in general relativity in a systematic and rigorous way by considering
a one-parameter family of metrics, $g_{ab} (\lambda)$, corresponding
to having a body (or black hole) that is ``scaled down'' to zero size
and mass in an appropriate manner. We prove that the limiting
worldline of such a one-parameter family must be a geodesic of the
background metric, $g_{ab} (\lambda=0)$. Gravitational self-force---as
well as the force due to coupling of the spin of the body to
curvature---then arises as a first-order perturbative correction in
$\lambda$ to this worldline. No assumptions are made in our analysis
apart from the smoothness and limit properties of the one-parameter
family of metrics, $g_{ab} (\lambda)$. Our approach should provide a
framework for systematically calculating higher order corrections to
gravitational self-force, including higher multipole effects, although
we do not attempt to go beyond first order calculations here. The
status of the MiSaTaQuWa equations is explained.

\end{abstract}

\maketitle 

\section{Introduction}

The physical content of general relativity is contained in Einstein's
equation, which has a well-posed initial value formulation (see, e.g.,
\cite{wald}).  In principle, therefore, to determine the
motion of bodies in general relativity---such as binary neutron stars
or black holes---one simply needs to provide appropriate initial data
(satisfying the constraint equations) on a spacelike slice and then
evolve this data via Einstein's equation. However, in practice, it is
generally impossible to find exact solutions of physical interest
describing the motion of bodies by analytic methods. Although it now
is possible to find solutions numerically in many cases of interest,
it is difficult and cumbersome to do so, and one may overlook subtle
effects and/or remain unenlightened about some basic general features
of the solutions. Therefore, it is of considerable interest to develop
methods that yield approximate descriptions of motion in some cases of
interest.

In general, the motion of a body of finite size will depend on the
details of its composition as well as the details of its internal
states of motion. Therefore, one can expect to get a simple
description of motion only in some kind of ``point particle
limit''. However, Einstein's equation is nonlinear, and a
straightforward analysis \cite{geroch-traschen} shows that it does not
make any mathematical sense to consider solutions of Einstein's
equation with a distributional stress-energy tensor supported on a
worldline\footnote{Nevertheless, action principles corresponding to general
  relativity with point particle sources are commonly written down
  (see, e.g., eqs.(12.1.6) and (12.4.1)-(12.4.2) of
  \cite{weinberg}).  There are no solutions to the equations of
  motion resulting from such action principles. By contrast,
  distributional solutions of Einstein's equation with support on a
  timelike hypersurface (i.e., ``shells'') do make mathematical sense
  \cite{israel,geroch-traschen}.}. Physically, if one tried to shrink a body down
to zero radius at fixed mass, collapse to a black hole would occur
before the point particle limit could be reached.

Distributional stress-energy tensors supported on a world-line
\textit{do} make mathematical sense in the context of the linearized
Einstein equation.  Therefore, one might begin a treatment of
gravitational self-force by considering a metric perturbation, $h_{ab}$, 
in a background metric, $g_{ab}$, sourced
by the stress-energy tensor of a ``point particle'' of mass M, given in coordinates $(t,x^i)$ by
\begin{equation}\label{eq:linpp}
G^{(1)}_{ab}[h](t,x^i) = 8 \pi M u_a(t) u_b(t) \frac{\delta^{(3)}(x^i
  - z^i(t))}{\sqrt{-g}} \frac{d\tau}{dt} ,
\end{equation}
where $u^a$ is the unit tangent (i.e., 4-velocity) of the worldline
$\gamma$ defined by $x^i(t) = z^i(t)$, and $\tau$ is the proper time along $\gamma$. (Here $\delta^{(3)}(x^i -
z^i(t))$ is the ``coordinate delta function'', i.e., $\int
\delta^{(3)}(x^i - z^i(t)) d^3x^i = 1$. The right side also could be
written covariantly as $8 \pi M \int_\gamma \delta_4(x,z(\tau)) u_a(\tau)
u_b(\tau) d\tau$ where $\delta_4$ is the covariant 4-dimensional
delta-function and $\tau$ denotes the proper time along $\gamma$.)
However, this approach presents two major difficulties. 

First, the
linearized Bianchi identity implies that the point particle
stress-energy must be conserved. However, as we shall see explicitly
in section \ref{sec:geodesic} below, this requires that the worldline
$\gamma$ of the particle be a geodesic of
the background spacetime. Therefore, there are no solutions to
equation \eqref{eq:linpp} for non-geodesic source curves, making it
hopeless to use this equation to derive
corrections to geodesic motion. This difficulty has been circumvented
in 
\cite{mino-sasaki-tanaka, quinn-wald, poisson, detweiler-whiting} 
and other references by modifying \eqref{eq:linpp} as follows. Choose the
Lorenz gauge condition, so that equation \eqref{eq:linpp}
takes the form
\begin{eqnarray}
\label{eq:wave}
\nabla^c \nabla_c \tilde{h}_{ab} - 2 R^c{}_{ab}{}^d \tilde{h}_{cd} &=&
- 16 \pi M u_a(t) u_b(t) \frac{\delta^{(3)}(x^i - z^i(t))}{\sqrt{-g}} \frac{d\tau}{dt},
\\ \nabla^b \tilde{h}_{ab} &=& 0.
\label{eq:gauge}
\end{eqnarray}
where $\tilde{h}_{ab} \equiv h_{ab} - \frac{1}{2} h g_{ab}$ with $h =
h_{ab} g^{ab}$. Equation \eqref{eq:wave} by itself has solutions for
any source curve $\gamma$; it is only when the Lorenz gauge condition
\eqref{eq:gauge} is adjoined that the equations are equivalent to the
linearized Einstein equation and geodesic motion is
enforced. Therefore, if one solves the Lorenz-gauge form \eqref{eq:wave}
of the linearized
Einstein equation while simply \textit{ignoring} the Lorenz gauge
condition\footnote{In some references, the failure to satisfy
  eq.~(\ref{eq:gauge}) truly is ignored in the sense that it is not
  even pointed out that one has modified the linearized Einstein
  equation, and no attempt is made to justify this modification.}
that was used to derive \eqref{eq:wave}, one
allows for the possibility non-geodesic motion. Of course, this
``gauge relaxation'' of the linearized Einstein equation produces an
equation inequivalent to the original.  However, because deviations
from geodesic motion are expected to be small, the Lorenz gauge
violation should likewise be small, and it thus has been argued
\cite{quinn-wald} that solutions to the two systems should agree to
sufficient accuracy.

The second difficulty is that the solutions to eq.~(\ref{eq:wave}) are
singular on the worldine of the particle. Therefore, naive attempts to
compute corrections to the motion due to $h_{ab}$---such as demanding
that the particle move on a geodesic of $g_{ab} + h_{ab}$---are virtually
certain to encounter severe mathematical difficulties, analogous to the
difficulties encountered in treatments of the electromagnetic
self-force problem.

Despite these difficulties, there is a general consensus 
that in the approximation that spin and
higher multipole moments may be neglected,
the motion of a sufficiently small body (with no ``incoming radiation'') 
should be described 
by self consistently solving eq.~(\ref{eq:wave}) via the retarded
solution together with 
\begin{equation}\label{eq:MiSaTaQuWa-intro}
u^b \nabla_b u^a = - \frac{1}{2} (g^{ab} + u^a u^b)(2\nabla_d
h_{bc}^{\tiny \textrm{tail}}-\nabla_b h_{cd}^{\tiny
\textrm{tail}})\big |_{z(\tau)}u^c u^d \,\, ,
\end{equation}
where
\begin{equation}
h_{ab}^{\tiny \textrm{tail}}(x) = M
\int_{-\infty}^{\tau_{\textrm{ret}}^-}\left(G^+_{a b a'
b'}-\frac{1}{2}g_{ab}G^{+ \ c}_{\ c \ a ' b
'}\right)\left(x,z(\tau ')\right)u^{a '}u^{b '} d\tau ' \,\, ,
\end{equation}
with $G^+_{a b a'b'}$ the retarded Green's function for eq.~(\ref{eq:wave}), normalized with a factor of $-16 \pi$, following \cite{quinn-wald}.  The symbol $\tau_{\textrm{ret}}^-$ indicates that the range of the integral extends just short of the retarded time $\tau_{\textrm{ret}}$, so that only the ``tail'' (i.e., interior of the light cone) portion of the Green's function is used (see, e.g., reference \cite{poisson} for details).
Equations (\ref{eq:wave}) and (\ref{eq:MiSaTaQuWa-intro}) are known as the
MiSaTaQuWa equations, and have been derived by a variety of
approaches. However, there are difficulties with all of these
approaches.

One approach \cite{mino-sasaki-tanaka} that has been taken is to
parallel the analysis of \cite{dirac,dewitt-brehme} in the electromagnetic
case and use
conservation of effective gravitational stress-energy to determine the
motion. However, this use of distributional sources at second-order
in perturbation theory results in infinities that must be
``regularized''. Although these regularization procedures are
relatively natural-looking, the mathematical status of such a
derivation is unclear.

Another approach \cite{quinn-wald} is to postulate certain properties
that equations of gravitational self-force should satisfy. This yields
a mathematically clean derivation of the self-force corrected equations
of motion. However,
as the authors of \cite{quinn-wald} emphasized, the motion of bodies
in general relativity is fully described by Einstein's equation
together with the field equations of the matter sources, so no
additional postulates should be needed to obtain an equation of
motion, beyond the ``small body'' assumption and other such
approximations.  The analysis given by \cite{quinn-wald} shows that
equation \eqref{eq:MiSaTaQuWa-intro} follows from certain plausible
assumptions. However, their derivation is thus only a plausibility
argument for equation \eqref{eq:MiSaTaQuWa-intro}. Similar remarks apply to a later derivation by Poisson \cite{poisson} that uses a Green's
function decomposition developed by Detweiler and Whiting
\cite{detweiler-whiting}.

A third approach, taken by Mino, Sasaki, and Tanaka
\cite{mino-sasaki-tanaka} and later Poisson \cite{poisson}---building
on previous work of Burke \cite{burke}, d'Eath \cite{d'eath}, Kates \cite{kates}, Thorne and Hartle \cite{thorne-hartle}, and others---involves the use of matched
astymptotic expansions.  Here one assumes a metric form in the ``near
zone''---where the metric is assumed to be that of the body, with a
small correction due to the background spacetime---and in the ``far
zone''---where the metric is assumed to be that of the background
spacetime, with a small correction due to the body.  One then assumes
that there is an overlap region of the body where both metric forms
apply, and matches the expressions.  The equations of motion of the
body then arise from the matching conditions.  However, as we shall
indicate below, in addition to the ``Lorenz gauge relaxation'', there
are a number of assumptions and steps in these derivations that have
not been adequately justified.

A more rigorous approach to deriving gravitational self-force is
suggested by the work of Geroch and Jang \cite{geroch-jang} and later
Geroch and Ehlers \cite{geroch-ehlers} on geodesic motion of small
bodies (see also \cite{stuart}).  In \cite{geroch-jang}, one considers a fixed
spacetime background metric $g_{ab}$ and considers a smooth
one-parameter family of stress-energy smooth tensors $T_{ab}(\lambda)$
that satisfy the dominant energy condition and have support on a world
tube. As the parameter goes to zero, the world tube shrinks to a
timelike curve. It is then proven that this timelike curve must be a
geodesic.  This result was generalized in \cite{geroch-ehlers} to
allow $g_{ab}$ to also vary with $\lambda$ so that Einstein's equation
is satisfied. Within the framework of \cite{geroch-ehlers}, it
therefore should be possible to derive perturbative corrections to
geodesic motion, including gravitational self-force.  However, the
conditions imposed in \cite{geroch-ehlers} in effect require the mass
of the body to go to zero faster than $\lambda^2$.  Consequently, in
this approach, a self-force correction like \eqref{eq:MiSaTaQuWa-intro} to
the motion of the body would arise at higher order than finite size
effects and possibly other effects that would depend on the
composition of the body. Thus, while the work of \cite{geroch-ehlers}
provides a rigorous derivation of geodesic motion of a ``small body''
to lowest order, it is not a promising approach to derive 
gravitatational self-force
corrections to geodesic motion.

In this paper, we shall take an approach similar in spirit to that of
\cite{geroch-ehlers}, but we will consider a different smooth,
one-parameter family of metrics $g_{ab} (\lambda)$, wherein, in
effect, we have a body (or black hole) present that scales to zero
size in a self-similar manner, with both the size and the mass of the
body being proportional to $\lambda$. In the limit as $\lambda
\rightarrow 0$, the body (or black hole) shrinks down to a worldline,
$\gamma$. As in \cite{geroch-jang} and \cite{geroch-ehlers}, we prove
that $\gamma$ must be a geodesic of the ``background spacetime''
$g_{ab} (\lambda = 0)$---although our method of proving this differs
significantly from \cite{geroch-jang} and \cite{geroch-ehlers}. To
first order in $\lambda$, the correction to the motion is described by
a vector field, $Z^i$, on $\gamma$, which gives the ``infinitesimal
displacement'' to the new worldline.  We will show that, for any such
one parameter family $g_{ab} (\lambda)$, in the Lorenz gauge
$Z^i(\tau)$ satisfies
\begin{equation}\label{eq:EOM-intro}
\frac{d^2Z^i}{dt^2} = \frac{1}{2M} S^{kl} {R_{kl0}}^i - {R_{0j0}}^iZ^j
- \left( {h^{\textrm{\tiny tail}}}^i{}_{0,0} - \frac{1}{2}
h^{\textrm{\tiny tail}}_{\ \ \ 00}{}^{,i} \right) \,\, .
\end{equation}

Here $M$ and $S_{ij}$ are, respectively, the mass and spin of the
body. The terms in parentheses on the right side of this equation correspond
exactly to the gravitational self-force term in
eq.\eqref{eq:MiSaTaQuWa-intro}; the first term is the Papapetrou
spin-force \cite{papapetrou}; the second term is simply the usual
right hand side of the geodesic deviation equation.
Equation \eqref{eq:EOM-intro} is ``universal'' in the
sense that it holds for any one-parameter family satisfying our
assumptions, so it holds equally well for a (sufficiently small)
blob of ordinary matter or a (sufficiently small) black hole.

Our derivation of \eqref{eq:EOM-intro} is closely related to the
matched asymptotic expansions analyses. However, our derivation is a
rigorous, perturbative result.  In addition, we eliminate a number of
ad hoc, unjustified, and/or unnecssary assumptions made in
previous approaches, including assumptions about the form of the
``body metric'' and its perturbations, assumptions about rate of
change of these quantities with time, the imposition of certain gauge
conditions, the imposition of boundary conditions at the body, and,
most importantly, the step of Lorenz gauge relaxation. Furthermore, in
our approach, the notion of a ``point particle'' is a concept that is
{\it derived} rather than assumed. It also will be manifest in our
approach that the results hold for all bodies (or black holes) and
that the physical results do not depend on a choice of gauge (although
$Z^i(\tau)$ itself is a gauge dependent quantity, i.e., the
description of the corrections to particle motion depend on how the
spacetimes at different $\lambda$ are identified---see section
\ref{sec:dipole} and Appendix A). In particular, because the Lorenz
gauge plays no preferred role in our derivation (aside from being a
computationally convenient choice), our gauge transformation law is
not, as in previous work \cite{barack-ori}, restricted to gauges
continuously related to the Lorenz gauge. Our approach also holds out
the possibility of being extended so as to systematically take higher
order corrections into account. However, we shall not attempt to
undertake such an extension in this paper.

Although \eqref{eq:EOM-intro} holds rigorously as a first-order
perturbative correction to geodesic motion, this equation would not be
expected to provide a good global-in-time approximation to motion,
since the small local corrections to the motion given by
\eqref{eq:EOM-intro} may accumulate with time, and
eq.~\eqref{eq:EOM-intro} would not be expected to provide a good
description of the perturbed motion when $Z^i$ becomes large. We will
argue in this paper that the MiSaTaQuWa equations,
eqs.~\eqref{eq:wave} and \eqref{eq:MiSaTaQuWa-intro}, should provide a
much better global-in-time approximation to motion than
eq.\eqref{eq:EOM-intro}, and they therefore should be used for
self-consistent calculations of the motion of a small body, such as
for calculations of extreme-mass-ratio inspiral\footnote{This viewpoint is in contrast to that of \cite{poisson}, where it is argued that second-order perturbations are needed for self-consistent evolution.}.

We note in passing that, in contrast to Einstein's equation, Maxwell's
equations are linear, and it makes perfectly good mathematical sense
to consider distributional solutions to Maxwell's equations with point
particle sources. However, if the charge-current sources are not
specified in advance but rather are determined by solving the matter
equations of motion---which are assumed to be such that the total stress-energy
of the matter and electromagnetic field is conserved---then the full,
coupled system of Maxwell's equations together with the equations of
motion of the sources becomes nonlinear in the the electromagnetic
field. Point particle sources do not make mathematical sense in this
context either. It is for this reason that---despite more than a
century of work on this problem---there is no mathematically
legitimate derivation of the electromagnetic self-force on a charged
particle. The methods of this paper can be used to rigorously derive a
perturbative expression for electromagnetic self-force by considering
suitable one-parameter families of coupled electromagnetic-matter
systems, and we shall carry out this analysis elsewhere \cite{gralla-wald}. However, we shall restrict consideration in this paper
to the gravitational case.

This paper is organized as follows. In section \ref{sec:example}, we
motivate the kind of one-parameter family of metrics, $g_{ab}
(\lambda)$, that we will consider by examining the one-parameter
family of Schwarzschild-deSitter spacetimes with black hole mass equal
to $\lambda$. One way of taking the limit as $\lambda \rightarrow 0$
yields deSitter spacetime. We refer to this way of taking the limit as
the ``ordinary limit''. But we show that if we take the limit as
$\lambda \rightarrow 0$ in another way and also rescale the metric by
$\lambda^{-2}$, we obtain Schwarzschild spacetime. We refer to this
second way of taking the limit as $\lambda \rightarrow 0$ as the
``scaled limit''. The scaled limit can be taken at any time $t_0$ on
the worldline $\gamma$.  The simultaneous existence of both types of
limits defines the kind of one parameter family of metrics we seek,
wherein a body (or black hole) is shrinking down to a world-line
$\gamma$ in an asymptotically self-similar manner. The precise,
general conditions we impose on $g_{ab} (\lambda)$ are formulated in
section \ref{sec:example}. Some basic properties of $g_{ab} (\lambda)$
that follow directly from our assumptions are given in section
\ref{sec:assumptions}. In particular, we obtain general ``far zone''
and ``near zone'' expansions and we show that, at any $t_0$, the
scaled limit always yields a stationary, asymptotically flat spacetime
at $\lambda = 0$. In section \ref{sec:geodesic}, we prove that
$\gamma$ must be a geodesic of the ``background spacetime'' (i.e., the
spacetime at $\lambda = 0$ resulting from taking the ordinary limit).
In other words, to zeroth order in $\lambda$ a small body or black
hole always moves on a geodesic. We also show that, to first order in
$\lambda$, the metric perturbation associated with such a body or
black hole corresponds to that of a structureless ``point particle''.
In section \ref{sec:dipole}, we define the motion of the body (or
black hole) to first order in $\lambda$ by finding a coordinate
displacement that makes the mass dipole moment of the stationary,
asymptotically flat spacetime appearing in the scaled limit vanish.
(This can be interpreted as a displacement to the ``center of mass''
of the body.)  In section \ref{sec:calculation}, we then derive
eq.\eqref{eq:EOM-intro} as the first order in $\lambda$ correction to
$\gamma$ in the Lorenz gauge. (An appendix provides the transformation
to other gauges.) Finally, in section \ref{sec:beyond} we explain
the status of the MiSaTaQuWa equation \eqref{eq:MiSaTaQuWa-intro}. Our
spacetime conventions are those of Wald \cite{wald}, and we work in
units where $G=c=1$. Lower case Latin indices early in the alphabet
($a,b,c,...$) will denote abstract spacetime indices; Greek indices
will denote coordinate components of these tensors; Latin indices from
mid-alphabet ($i,j,k,...$) will denote spatial coordinate components.

\section{Motivating Example and Assumptions}\label{sec:example}

As discussed in the introduction, we seek a one-parameter family of
metrics $g_{ab}(\lambda)$ that describes a material body or black hole
that ``shrinks down to zero size'' in an asymptotically self-similar
manner. In order to motivate the general conditions on
$g_{ab}(\lambda)$ that we shall impose, we
consider here an extremely simple example of the type of one-parameter
family that we seek. This example will provide an illustration of the
two types of limits that we shall use to characterize $g_{ab}(\lambda)$.

Our example is built from Schwarzschild-deSitter spacetime,
\begin{equation}
ds^2 = - \left( 1 - \frac{2 M_0}{r} - C_0 r^2 \right) dt^2 + 
\left( 1 - \frac{2 M_0}{r} - C_0 r^2 \right)^{-1} dr^2 + r^2 d \Omega ^2 \,\, .
\end{equation}
(This metric, of course, is a solution to the vacuum Einstein's
equation with a cosmological constant $\Lambda = \frac{2}{3} C_0$
rather than a solution with $\Lambda = 0$, but the field equations
will not play any role in the considerations of this section; we
prefer to use this example because of its simplicity and familiarity.)
The desired one-parameter family is
\begin{equation}\label{eq:family}
ds^2(\lambda) = - \left( 1 - \frac{2 M_0 \lambda}{r} - C_0 r^2 \right) dt^2 + \left( 1 - \frac{2 M_0 \lambda}{r} - C_0 r^2 \right)^{-1} dr^2 + r^2 d \Omega ^2,
\end{equation}
where we consider only the portion of the spacetime with $r > \lambda
R_0$ for some $R_0 > 2 M_0$. For each $\lambda$, this spacetime
describes the exterior gravitational field of a spherical body or
black hole of mass $\lambda M_0$ in an asymptotically deSitter
spacetime. As $\lambda \rightarrow 0$, the body/black hole shrinks to
zero size and mass. For $\lambda = 0$, the spacetime is deSitter
spacetime, which extends smoothly to the worldline $r=0$,
corresponding to where the body/black hole was just before it
``disappeared''.

As explained clearly in \cite{geroch-limits}, the limit of a one-parameter family of metrics $g_{ab}(\lambda)$ depends on how the spacetime manifolds at
different values of $\lambda$ are identified. This identification of
spacetime manifolds at different $\lambda$ can be specified by
choosing coordinates $x^\mu$ for each $\lambda$ and identifying points
labeled by the same value of the coordinates $x^\mu$. If we use the
coordinates $(t,r,\theta, \phi)$ in which the one-parameter family of
metrics \eqref{eq:family} was presented to do the identification, then
it is obvious that the limit as $\lambda \rightarrow 0$ of
$g_{ab}(\lambda)$ is the deSitter metric
\begin{equation}\label{eq:deSitter}
ds^2(\lambda=0) = - \left( 1 - C_0 r^2 \right) dt^2 +
\left( 1 - C_0 r^2 \right)^{-1} dr^2 + r^2 d \Omega ^2.
\end{equation}
This corresponds to the view that the body/black hole shrinks to zero
size and mass and ``disappears'' as $\lambda \rightarrow 0$.

However, there is another way of taking the limit of $g_{ab}(\lambda)$
as $\lambda \rightarrow 0$; the existence of this second limit is one
of the key ideas in this paper. Choose an arbitrary time $t_0$ and,
for $\lambda > 0$, introduce new time and radial coordinates by
$\bar{t} \equiv (t-t_0)/\lambda$ and $\bar{r} \equiv r/\lambda$. In
the new coordinates, the one-parameter family of metrics becomes
\begin{equation}
ds^2(\lambda) = - \lambda^2 \left( 1 - \frac{2 M_0}{\bar{r}} - C_0 \lambda^2 \bar{r}^2 \right) d\bar{t}^2 + \lambda^2 \left( 1 - \frac{2 M_0}{\bar{r}} - C_0 \lambda^2 \bar{r}^2 \right)^{-1} d\bar{r}^2 + \lambda^2 \bar{r}^2 d \Omega ^2, \bar{r} > R_0.
\label{gbar}
\end{equation}
We now consider the limit as $\lambda \rightarrow 0$ by identifying
the spacetimes with different $\lambda$ at the same values of the
barred coordinates. It is clear by inspection of eq.(\ref{gbar}) that
the the limit of $g_{ab}(\lambda)$ as $\lambda \rightarrow 0$ at fixed
$\bar{x}^\mu$ exists, but is zero. In essence, the spacetime interval
between any two events labeled by $\bar{x}_1^\mu$ and $\bar{x}_2^\mu$
goes to zero as $\lambda \rightarrow 0$ because these events are
converging to the same point on $\gamma$. Thus, this limit of
$g_{ab}(\lambda)$ exists but is not very interesting.  However, an
interesting limit can be taken by considering a new one-parameter
family of metrics $\bar{g}_{ab}(\lambda)$ by\footnote{A scaling of
  space (but not time) has previously been considered by Futamase
  \cite{futamase} in the post-Newtonian context.  Scaled coordinates also appear in the work of D'Eath \cite{d'eath} and Kates \cite{kates}.}
\begin{equation}
\bar{g}_{\bar{\mu} \bar{\nu}} \equiv \lambda^{-2} g_{\bar{\mu} \bar{\nu}},
\end{equation}
so that
\begin{equation}\label{eq:barred}
d\bar{s}^2(\lambda) = \left( 1 - \frac{2 M_0}{\bar{r}} - C_0 \lambda^2 \bar{r}^2 \right) d\bar{t}^2 + \left( 1 - \frac{2 M_0}{\bar{r}} - C_0 \lambda^2 \bar{r}^2 \right)^{-1} d\bar{r}^2 + \bar{r}^2 d \Omega ^2, \bar{r} > R_0,
\end{equation}
By inspection, the limit of this family of metrics is simply,
\begin{equation}\label{eq:body}
d\bar{s}^2|_{\lambda = 0} = - \left( 1 - \frac{2 M_0}{\bar{r}} \right) d\bar{t}^2 + \left( 1 - \frac{2 M_0}{\bar{r}} \right)^{-1} d\bar{r}^2 + \bar{r}^2 d \Omega ^2, \bar{r} > R_0 \,\, ,
\end{equation}
which is just the Schwarzschild metric with mass $M_0$.  

The meaning of this second limit can be understood as follows. As
already discussed above, the one-parameter family of metrics
(\ref{eq:family}) describes the exterior gravitational field of a
spherical body or black hole that shrinks to zero size and mass as
$\lambda \rightarrow 0$. The second limit corresponds to how this
family of spacetimes looks to the family of observers placed at the
events labeled by fixed values of $\bar{x}^\mu$. In going from, say,
the $\lambda = 1$ to the $\lambda = 1/100$ spacetime, each observer
will see that the body/black hole has shrunk in size and mass by a
factor of $100$ and each observer also will find himself ``closer to
the origin'' by this same factor of $100$. Suppose now that this
family of observers also ``changes units'' by this same factor of
$100$, i.e., they use centimeters rather than meters to measure
distances. Then, except for small effects due to the deSitter
background, the family of observers for the $\lambda = 1/100$
spacetime will report the same results (expressed in centimeters) as
the observers for the $\lambda = 1$ spacetime had reported (in
meters). In the limit as $\lambda \rightarrow 0$, these observers
simply see a Schwarzschild black hole of mass $M_0$, since the effects
of the deSitter background on what these observers will report 
disappear entirely in this limit.

\begin{figure}
\includegraphics[width=3.00in]{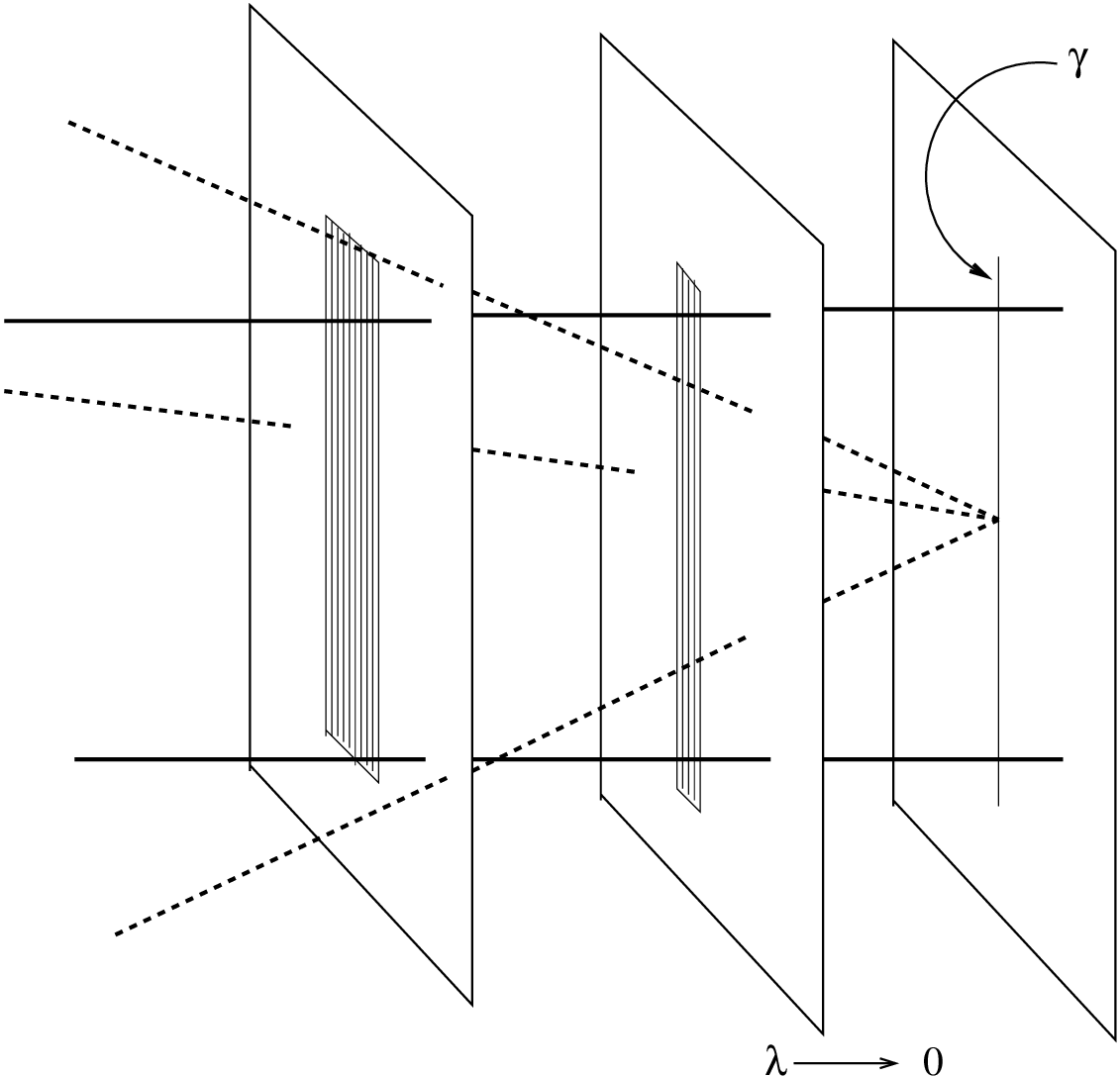}
\caption{A spacetime diagram illustrating the type of one-parameter
  family we wish to consider, as well as the two limits we define.  As
  $\lambda \rightarrow 0$, the body shrinks and finally disappears,
  leaving behind a smooth background spacetime with a preferred
  world-line, $\gamma$, picked out.  The solid lines illustrate this
  ``ordinary limit'' of $\lambda \rightarrow 0$ at fixed $r$, which is
  taken along paths that terminate away from $\gamma$ (i.e., $r>0$).
  By contrast, the ``scaled limit'' as $\lambda \rightarrow 0$, shown
  in dashed lines, is taken along paths at fixed $\bar{r}$ that
  converge to a point on $\gamma$.}
\label{fig:2limits-spacetime}
\end{figure}

We will refer to the first type of limit (i.e., the limit 
as $\lambda \rightarrow 0$
of $g_{ab}(\lambda)$ taken at fixed
$x^\mu$) as the {\it ordinary limit} of $g_{ab}(\lambda)$, and we will
refer to the second limit (i.e., the limit as $\lambda \rightarrow 0$
of $\lambda^{-2} g_{ab}(\lambda)$ taken at fixed
$\bar{x}^\mu$) as the {\it scaled limit} of $g_{ab}(\lambda)$.
The simultaneous existence of both types of the above limits contains
a great deal of relevant information about the one-parameter family of
spacetimes (\ref{eq:family}). In essence, the existence of the first
type of limit is telling us that the body/black hole is shrinking down
to a worldline $\gamma$, with its mass going to zero (at least) as
rapidly as its radius. The existence of the second type of limit is
telling us that the body/black hole is shrinking to zero size in an
asymptotically self-similar manner: In particular, its mass is
proportional to its size, its shape is not changing, and it is not
undergoing any (high frequency) oscillations in time.
Figure \ref{fig:2limits-spacetime} illustrates the two limits we
consider.

We wish now to abstract from the above example the general conditions
to impose upon one-parameter families that express in a simple and
precise way the condition that we have an arbitrary body (or black
hole) that is shrinking to zero size---in an asymptotically
self-similar way---in an arbitrary background spacetime. Most of the
remainder of this paper will be devoted to obtaining ``equations of
motion'' for these bodies that are accurate enough to include
gravitational self-force effects. A first attempt at specifying the
type of one-parameter families $g_{ab}(\lambda)$ that we seek is as follows:

\begin{itemize}

\item 
(i) {\it Existence of the ``ordinary limit''}: $g_{ab}(\lambda)$ is such
  that there exists coordinates $x^\alpha$ such that $g_{\mu
    \nu}(\lambda, x^\alpha)$ is jointly smooth in $(\lambda,
  x^\alpha)$, at least for $r > \bar{R} \lambda$ for some constant
  $\bar{R}$, where $r \equiv \sqrt{\sum (x^i)^2}$ ($i=1,2,3$). For all
  $\lambda$ and for $r > \bar{R} \lambda$, $g_{ab}(\lambda)$ is a
  vacuum solution of Einstein's equation. Furthermore, $g_{\mu
    \nu}(\lambda = 0, x^\alpha)$ is smooth in $x^\alpha$, including at
  $r= 0$, and, for $\lambda=0$, the curve $\gamma$ defined by $r= 0$
  is timelike.

\item
(ii) {\it Existence of the ``scaled limit''}: For each $t_0$, we define
  $\bar{t} \equiv (t-t_0)/\lambda$, $\bar{x}^i \equiv x^i/\lambda$.
  Then the metric $\bar{g}_{\bar{\mu} \bar{\nu}}(\lambda; t_0;
  \bar{x}^\alpha) \equiv \lambda^{-2} g_{\bar{\mu} \bar{\nu}}(\lambda;
  t_0; \bar{x}^\alpha)$ is jointly smooth in $(\lambda, t_0;
  \bar{x}^\alpha)$ for $\bar{r} \equiv r/\lambda > \bar{R}$.

\end{itemize}

Here we have used the notation $g_{\bar{\mu} \bar{\nu}}$ to denote the
components of $g_{ab}$ in the $\bar{x}^\alpha$ coordinates, whereas
$\bar{g}_{\bar{\mu} \bar{\nu}}$ denotes the components of
$\bar{g}_{ab}$ in the $\bar{x}^\alpha$ coordinates. It should be noted
that, since the barred coordinates differ only by scale (and shift of
time origin) from the corresponding unbarred coordinates, we
have\footnote{Note that in this equation and in other equations
  occurring later in this paper, we relate components of tensors in
  the barred coordinates to the corresponding components of tensors in
  unbarred coordinates. Thus, a bar appears over the indices on the
  left side of this equation, but not over the indices appearing on
  the right side of this equation.}
\begin{equation}
 g_{\bar{\mu} \bar{\nu}} = \lambda^2 g_{\mu \nu} \,\, .
\end{equation}
Consequently, we have
\begin{equation}
\bar{g}_{\bar{\mu} \bar{\nu}}(\lambda; t_0; \bar{t},\bar{x}^i)
= g_{\mu \nu} (\lambda; t_0 + \lambda \bar{t}, \lambda \bar{x}^i)
\label{barg}
\end{equation}
since there is a cancelation of the
factors of $\lambda$ resulting from the definition of $\bar{g}_{ab}$ 
and the coordinate rescalings. It also should be noted that there
is a redundancy in our description of the one-parameter family of metrics
when taking the scaled limit: We define a scaled limit for all values 
of $t_0$, but these arise from a single one-parameter family of metrics
$g_{ab}(\lambda)$. Indeed, it is not difficult to see that we have
\begin{equation}
\bar{g}_{\bar{\mu} \bar{\nu}}(\lambda; t_0; \bar{t} + \bar{s},\bar{x}^i)
= \bar{g}_{\bar{\mu} \bar{\nu}}(\lambda; t_0 + \lambda \bar{s}; \bar{t},\bar{x}^i)
\,\, .
\label{consist}
\end{equation}

\begin{figure}
  \subfigure[The two limits in terms of $r$ and $\lambda$.  A
  constant-$\lambda$ spacetime is shown as a thick line.  The shaded
  region corresponds to the interior of the (shrinking) body, about
  which we make no assumptions.  The ordinary limit is represented by
  solid lines and the scaled limit is represented by dashed lines.
  While the ordinary background metric is on the $r$-axis, the scaled
  background metric is contained in the discontinuous behavior of the
  metric family at
  $r=\lambda=0$.]{\label{fig:2limits}\includegraphics[width=3.00in]{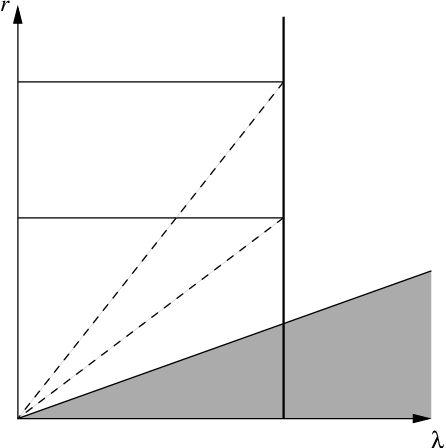}}
  \subfigure[The two limits in terms of $\alpha$ and $\beta$.  In the
  new variables, the two types of limit appear on a more equal
  footing, with the ordinary and scaled background metrics placed on
  either axis.  The body is ``pushed out'' to finite $\beta$, so
  that assumptions made in a neighborhood of $\alpha=\beta=0$ make no
  reference to the body.]
  {\label{fig:2limits-ab}\includegraphics[width=3.00in]{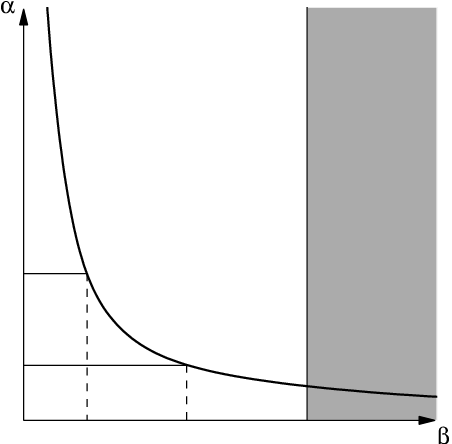}}
\caption{The two limits}
\end{figure}

In fact, our requirements on $g_{ab}$ of the existence of both an
``ordinary limit'' and a ``scaled limit'' are not quite strong enough
to properly specify the one-parameter families we seek. To explain
this and obtain the desired strengthened condition, it is convenient
to define the new variables
\begin{equation}
\alpha \equiv r \, , \,\,\,\,\,\,\,\,\,\, \beta \equiv \lambda/r \,\, ,
\label{alphabeta}
\end{equation}
where the range of $\beta$ is $0 \leq \beta < 1/\bar{R}$.
Let $f$ denote a component of $g_{ab}(\lambda)$ in the coordinates
$x^\alpha$. However, instead of viewing $f$ as a function of
$(\lambda, x^\alpha)$, we view $f$ as a function of $(\alpha, \beta,
t, \theta, \phi)$, where $\theta$ and $\phi$ are defined in terms of
$x^i$ by the usual formula for spherical polar angles. In terms of
these new variables, taking the ``ordinary limit'' corresponds to
letting $\beta \rightarrow 0$ at any fixed $\alpha > 0$, whereas taking
the ``scaled limit'' corresponds to letting $\alpha \rightarrow 0$ at
any fixed $\beta > 0$ (see figure \ref{fig:2limits-ab}). Now, our assumptions concerning the ordinary
limit imply that, at fixed $(t,\theta, \phi)$ and at fixed $\alpha >
0$, $f$ depends smoothly on $\beta$, including at $\beta = 0$. On the
other hand, our assumptions concerning the scaled limit imply that
at fixed $(t,\theta, \phi)$ and at fixed $\beta > 0$, $f$ is smooth in
$\alpha$. Furthermore, the last condition in the ordinary limit
implies that for $\beta = 0$ and fixed $(t,\theta, \phi)$, $f$ is smooth
in $\alpha$, including at $\alpha = 0$.

Thus, at fixed $(t,\theta, \phi)$, our previously stated assumptions
imply that $f$ is well defined at the ``origin'' $(\alpha, \beta) =
(0,0)$, and is smooth in $\alpha$ along the $\alpha$-axis (i.e.,
$\beta = 0$). However, our previously stated assumptions do not say
anything about the continuity or smoothness of $f$ as $(\alpha, \beta)
\rightarrow (0,0)$ from directions other than along the
$\alpha$-axis. Such limiting behavior would correspond to letting $r
\rightarrow 0$ as $\lambda \rightarrow 0$ but at a rate {\it slower}
than $\lambda$, i.e., such that $r/\lambda \rightarrow \infty$. To see
the meaning and relevance of this limiting behavior, let us return to
our original motivating example, eq. (\ref{eq:family}) and take $f$ to
be the time-time component of this metric\footnote{Note that if we
  wished to consider other components of this metric, we would have to
  transform back from the coordinates $(r,\theta,\phi)$ to
  ``Cartesian-like'' coordinates $x^i$ that are well behaved at the
  origin $x^i = 0$ when $\lambda = 0$}. In terms of our new variables
(\ref{alphabeta}), we have
\begin{equation}
f = -(1 - 2 M_0 \beta - C_0 \alpha^2) \,\, ,
\end{equation}
which is jointly smooth in $(\alpha, \beta)$ at $(0,0)$. By contrast,
suppose we had a one-parameter family of metrics $\tilde{g}_{ab}
(\lambda)$ that satisfies our above assumptions about the ordinary and
scaled limits, but fails to be jointly smooth in $(\alpha, \beta)$ at
$(0,0)$. For example, suppose the time-time 
component of such a one-parameter family
varied as
\begin{equation}
\tilde{f} = -\left(1 - \frac{\alpha \beta}{\alpha^2 + \beta^2}\right) \,\, .
\end{equation}
In terms of the original variables $(\lambda, r)$, this corresponds to
behavior of the form
\begin{equation}
\tilde{f} = - \left(1 - \frac{\lambda r^2}{\lambda^2 + r^4}\right) \,\, .
\end{equation}
If we take the ``ordinary limit'' ($\lambda \rightarrow 0$ at fixed
$r>0$) of $\tilde{f}$, we find that $\tilde{f}$ smoothly goes to
$-1$. Similarly, if we take the ``scaled limit'' ($\lambda \rightarrow
0$ at fixed $\bar{r} = r/\lambda > 0$), we also find that $\tilde{f}$
smoothly goes to $-1$.  However, suppose we let $\lambda \rightarrow 0$
but let $r$ go to zero as $r = c \sqrt{\lambda}$. Then $\tilde{f}$
will approach a different limit, namely $c^2/(1+ c^4)$.  In essence,
$\tilde{g}_{ab} (\lambda)$ corresponds to a one-parameter family in
which there is a ``bump of curvature'' at $r \propto
\sqrt{\lambda}$. Although this ``bump of curvature'' does not register
when one takes the ordinary or scaled limits, it is present in the
one-parameter family of spacetimes and represents unacceptable limiting
behavior as $\lambda \rightarrow 0$ of this one-parameter family.

In order to eliminate this kind of non-uniform behavior in $\lambda$ and $r$,
we now impose the following additional condition:

\begin{itemize}
\item
(iii) {\it Uniformity condition}: Each component of $g_{ab}(\lambda)$ in the coordinates $x^\mu$ is a
jointly smooth function of the variables $(\alpha,\beta)$ at $(0,0)$ (at fixed ($t,\theta,\phi$)), where $\alpha$ and $\beta$ are defined
by eq.(\ref{alphabeta}).  [Note added: Joint smoothness in all variables is actually required, so that our series expansions in $\alpha$ and $\beta$ produce smooth coefficients.]
\end{itemize}

Assumptions (i)-(iii) constitute all of the conditions that we shall impose
on $g_{ab}(\lambda)$. No additional assumptions will be made in this paper.

We note that the coordinate freedom allowed by our conditions are
precisely all coordinate transformations
\begin{equation}\label{eq:mushi-mushi}
x^\mu \rightarrow x'^\mu(\lambda, x^\nu)
\end{equation}
such that $x'^\mu(\lambda, x^\nu)$ is jointly smooth in $(\lambda,
x^\nu)$ for all $r > C \lambda$ for some constant $C$, and such that
the Jacobian matrix $\partial x'^\mu/\partial x^\nu$ is jointly smooth
in $(\alpha, \beta)$ at $(0,0)$ at fixed $(t, \theta, \phi)$. [Note added: Please see arXiv:1104.5205.]

It should be emphasized that our assumtions place absolutely no
restrictions on the one-parameter family of spacetimes for $r <
\lambda \bar{R}$, i.e., this portion of these spacetimes could equally
well be ``filled in'' with ordinary matter or a black
hole\footnote{Indeed, it could also be ``filled in'' with ``exotic
  matter'' (failing to satisfy, say, the dominant energy condition) or
  a naked singularity (of positive or negative mass). However, in
  cases where there fails to be a well posed initial value formulation
  (as would occur with certain types of exotic matter and with naked
  singularities) and/or there exist instabilities (as would occur with
  other types of exotic matter), if is far from clear that one should
  expect there to exist a one-parameter family of solutions
  $g_{ab}(\lambda)$ satisfying our assumptions.}.  It also should be
noted that the ``large $r$'' region of the spacetime will not be
relevant to any of our considerations, so it is only necessary that
our conditions hold for $r < K$ for some constant $K$.

Finally, although it may not be obvious upon first reading, we note
that our assumptions concerning $g_{ab} (\lambda)$ are closely related
to the assumptions made in matched asymptotic expansion analyses. As
we shall see in the next section, in essence, our assumption about the
existence of an ordinary limit of $g_{ab} (\lambda)$ corresponds to
assuming the existence of a ``far zone'' expansion; our assumption
about the existence of a scaled limit of $g_{ab} (\lambda)$
corresponds to assuming the existence of a ``near zone'' expansion;
and our uniformity assumption corresponds closely to the assumption of
the existence of a ``buffer zone'' where both expansions are valid.

\section{Consequences of Our Assumptions}\label{sec:assumptions}

In this section, we derive some immediate consequences of the
assumptions of the previous section that will play a key role in our
analysis. These results will follow directly from the ``uniformity
condition'' and the consistency relation (\ref{consist}).

Since, by the uniformity assumption, the coordinate components of the
one-parameter family of metrics $g_{ab}(\lambda)$ are jointly smooth
in the variables $(\alpha, \beta)$ at $(0,0)$, we may approximate $g_{\mu \nu}$
by a joint Taylor series in $\alpha$ and $\beta$ to any finite orders
$N$ and $M$ by
\begin{equation}
g_{\mu \nu}(\lambda;t,r,\theta,\phi) = \displaystyle \sum_{n=0}^{N}
\sum_{m=0}^{M} \alpha^n \beta^m
(a_{\mu \nu})_{nm}(t,\theta,\phi) + O(\alpha^{N+1}) + O(\beta^{M+1}) \,\, .
\end{equation}  
Substituting for $\alpha$ and $\beta$ from eq.(\ref{alphabeta}), we have
\begin{equation}\label{eq:joint-taylor}
g_{\mu \nu}(\lambda;t,r,\theta,\phi) = \displaystyle \sum_{n=0}^{N}
\sum_{m=0}^{M} r^n \left(\frac{\lambda}{r}\right)^m
(a_{\mu \nu})_{nm}(t,\theta,\phi) \,\, ,
\end{equation}  
where here and in the following, we drop the error term.
We can rewrite this equation as a perturbation expansion in $\lambda$:
\begin{equation}\label{eq:rseries}
g_{\mu \nu}(\lambda;t,r,\theta,\phi) = \displaystyle \sum_{m=0}^{M} \lambda^m \sum_{n=0}^{N}r^{n-m} (a_{\mu \nu})_{nm}(t,\theta,\phi) \,\, .
\end{equation}
We will refer to eq.(\ref{eq:rseries}) as the {\it far zone expansion}
of $g_{ab}(\lambda)$. It should be noted that the $m$th-order term in
$\lambda$ in the far zone perturbation expansion has leading order
behavior of $1/r^m$ at small $r$. However, arbitrarily high positive
powers of $r$ may occur at each order in $\lambda$. Finally, we note
that the angular dependence of $(a_{\mu \nu})_{n0}(t,\theta,\phi)$ is
further restricted by the requirement that the metric components
$g_{\mu \nu}(\lambda = 0)$ must be smooth at $r=0$ when re-expressed
as functions of $x^i$. In particular, this implies that $(a_{\mu \nu})_{00}$
cannot have any angular dependence.

Equivalently, in view of eq.(\ref{barg}), we can expand 
$\bar{g}_{\bar{\mu} \bar{\nu}}$ as
\begin{eqnarray}
\bar{g}_{\bar{\mu} \bar{\nu}}(\lambda; t_0; \bar{t},\bar{r},\theta,\phi) &=&
\displaystyle \sum_{n=0}^{N} \sum_{m=0}^{M} (\lambda \bar{r})^n \left(
\frac{1}{\bar{r}} \right)^m (a_{\mu \nu})_{nm}(t_0 + \lambda \bar{t},\theta,\phi) \,\,  \nonumber \\
&=& \displaystyle \sum_{n=0}^{N} \sum_{m=0}^{M} \lambda^n \left(
\frac{1}{\bar{r}} \right)^{m-n} (a_{\mu \nu})_{nm}(t_0 + \lambda
\bar{t},\theta,\phi) \,\, .
\label{eq:rbarseries2}
\end{eqnarray}
By further expanding $(a_{\mu \nu})_{nm}$ in $\bar{t}$ about
$\bar{t}=0$, we obtain
\begin{equation}\label{eq:rbarseries3}
\bar{g}_{\bar{\mu} \bar{\nu}}(\lambda; t_0; \bar{t},\bar{r},\theta,\phi) =
\displaystyle \sum_{n=0}^{N} \sum_{m=0}^{M}\sum_{p=0}^{P}\lambda^{n+p} \bar{t}^p
\left(\frac{1}{\bar{r}} \right)^{m-n} (b_{\mu \nu})_{nmp}(t_0,\theta,\phi) \,\,,
\end{equation}
where 
\begin{equation}
(b_{\mu \nu})_{nmp} \equiv \frac{1}{p!} \frac{\partial^p}{\partial t^p} (a_{\mu \nu})_{nm}\bigg|_{t=t_0} \,\, . 
\label{bnmp}
\end{equation}
We can
rewrite this as a perturbation series expansion in $\lambda$:
\begin{equation}\label{eq:fullnearzone}
\bar{g}_{\bar{\mu} \bar{\nu}}(\lambda;t_0;\bar{t},\bar{r},\theta,\phi) =
\displaystyle \sum_{q=0}^{N+P} \lambda^q \sum_{p=0}^{{\rm min}(q,P)}
\sum_{m=0}^{M} \bar{t}^p \left( \frac{1}{\bar{r}} \right)^{m-q+p}
(b_{\mu \nu})_{(q-p)mp}(t_0;\theta,\phi) \,\, .
\end{equation}
We will refer to eq.(\ref{eq:fullnearzone}) as the {\it near zone
  expansion} of $g_{ab}(\lambda)$. We see from this formula that the
scaled metric, viewed as a perturbation series in $\lambda$, follows
the rule that the combined powers of $\bar{t}$ and $\bar{r}$ are
allowed to be only as positive as the order in perturbation theory. By
contrast inverse powers of $\bar{r}$ of arbitrarily high order are
always allowed. Of course, only non-negative powers of $\bar{t}$ can occur.

By inspection of eq.(\ref{eq:fullnearzone}), we see that the 
``background'' ($\lambda=0$) scaled metric is given by
\begin{equation}
\label{eq:fullnearzone2}
\bar{g}_{\bar{\mu} \bar{\nu}}(\lambda=0;t_0;\bar{t},\bar{r},\theta,\phi) =
\displaystyle \sum_{m=0}^{M} \left( \frac{1}{\bar{r}} \right)^m
(a_{\mu \nu})_{0m}(t_0;\theta,\phi) \,\, ,
\end{equation}
where we have used the fact that $(b_{\mu \nu})_{0m0} = (a_{\mu
  \nu})_{0m}$.  Thus, we see that there is no dependence of
$\bar{g}_{\bar{\mu} \bar{\nu}}(\lambda=0;t_0)$ on $\bar{t}$ and only non-positive
powers of $\bar{r}$ occur. {\it Thus, we see that $\bar{g}_{ab}(\lambda=0)$
is a stationary, asymptotically flat spacetime.} However, the limiting,
stationary, asymtotically flat spacetime that we obtain may depend on
the choice of the time, $t_0$, on the worldline, $\gamma$, about which
the scaling is done.

Our ``far zone expansion'', eq.(\ref{eq:rseries}), appears to
correspond closely to the far zone expansion used in matched
asymptotic expansion analyses \cite{mino-sasaki-tanaka,poisson}.
However, our ``near zone expansion'' differs in that we define a
separate expansion for each time $t_0$ rather than attempting a
uniform in time approximation with a single expansion.  Such
expansions require an additional ``quasi-static'' or slow-time
variation assumption for the evolution of the metric perturbations.
A further difference
is that the conclusion that the background ($\lambda = 0$) metric is
stationary and asymptotically flat has been derived here rather than
assumed. Indeed, in other analyses, a particular form of the
background metric (such as the Schwarzschild metric) is assumed, and
the possibility that this metric form might change with time (i.e.,
depend upon $t_0$) is not considered. In addition, in other analyses boundary conditions
at small $\bar{r}$ (such as regularity on the Schwarzschild horizon)
are imposed.  In our analysis, we make no
assumptions other than the assumptions (i)-(iii) stated in the
previous section. In particular, since we make no assumptions about
the form of the metric for $\bar{r} < \bar{R}$, we do not impose any
boundary conditions at small $\bar{r}$.

Finally, it
is also useful to express the consistency relation (\ref{consist})
in a simple,
differential form. We define
\begin{equation}\label{eq:consistency-deriv}
(K_{\mu \nu})_{npm}(\lambda;t_0;\bar{t},\bar{x}^i) \equiv \left(
  \frac{\partial}{\partial \lambda}\right)^n \left(
  \frac{\partial}{\partial t_0}\right)^p \left(
  \frac{\partial}{\partial \bar{t}}\right)^m
  \bar{g}(\lambda;t_0;\bar{t},\bar{x}^i)|_{\lambda=0, t_0, \bar{t}=0}
\end{equation}
Then, a short calculation shows that
\begin{equation}\label{eq:consistency}
K_{(n+1)p(m+1)} = (n+1) K_{n(p+1)m} \,\,
\end{equation}
as well as
\begin{equation}
\label{eq:consistency-1}
K_{npm} = 0 \textrm{ if } n<m \,\, .
\end{equation}
Setting $n=0$, we see that the last relation implies that
$\bar{g}(\lambda=0;t_0;\bar{t},\bar{x}^i)$ is stationary, as we have
already noted.

\section{Geodesic Motion} \label{sec:geodesic}

In this section, we will prove that the worldline $\gamma$ appearing
in assumption (i) of section \ref{sec:example} must, in fact, be a
geodesic of the background metric $g_{ab}(\lambda=0)$. This can be
interpreted as establishing that, to zeroth order in $\lambda$, any
body (or black hole) moves on a geodesic of the background
spacetime. In fact, we will show considerably more than this: We will
show that, to first order in $\lambda$, the far zone description of
$g_{ab}(\lambda)$ is that of a ``point particle''.  As previously
mentioned in the Introduction, our derivation of geodesic motion is
similar in spirit to that of \cite{geroch-ehlers} in that we consider
one-parameter families of solutions with a worldtube that shrinks down
to a curve $\gamma$, but the nature of the one-parameter families that
we consider here are very different from those considered by
\cite{geroch-ehlers}, and  our proof of geodesic motion is very
different as well.  Our derivation of geodesic motion also appears to differ significantly from pervious derivations using matched asymptotic expansions \cite{d'eath,kates,mino-sasaki-tanaka,poisson}.

We begin by writing the lowest order terms in the far zone expansion,
eq.~(\ref{eq:rseries}), as follows:
\begin{eqnarray}
g_{\alpha \beta}(\lambda) & = & (a_{\alpha \beta})_{00}(t) +
(a_{\alpha \beta})_{10} (t,\theta,\phi) r + O(r^2) \nonumber \\ & + & \lambda
\left[(a_{\alpha \beta})_{01} (t,\theta,\phi) \frac{1}{r} + 
(a_{\alpha \beta})_{11}(t,\theta,\phi) + O(r) \right] + O(\lambda^2)\, , r>0 \,\, ,
\end{eqnarray}
where we have used the fact that $(a_{\alpha \beta})_{00}$ can depend
only on $t$, as noted in the previous section. Since the worldline
$\gamma$, given by $x^i = 0$, was assumed to be timelike\footnote{We
  made this assumption explicitly in condition (i) of section
  \ref{sec:example}. However, if, instead, we had assumed that the
  ``interior region'' $r \leq \lambda \bar{R}$, were ``filled in''
  with matter satisfying the dominant energy condition, then it should
  be possible to prove that $\gamma$ must be timelike.} in the
spacetime $g_{ab}(\lambda = 0)$, without loss of generality, we may
choose our coordinates $x^\alpha$ so that $g_{\alpha
  \beta}(\lambda=0,x^i=0) = (a_{\alpha \beta})_{00}(t) =
\eta_{\alpha \beta}$. (One such possible choice of coordinates would
be Fermi normal coordinates with respect to $\gamma$ in the metric
$g_{ab}(\lambda = 0)$. We emphasize that we make the coordinate choice
$(a_{\alpha \beta})_{00} = \eta_{\alpha \beta}$ purely for
convenience---so that, e.g., coordinate time coincides with proper
time on $\gamma$---but it plays no essential role in our arguments.)
Choosing these coordinates, and letting $h_{\alpha \beta}$ denote the
$O(\lambda)$ piece of the metric, we see that $g_{\alpha
  \beta}(\lambda)$ takes the form
\begin{equation}
g_{\alpha \beta} = \eta_{\alpha \beta} + O(r) + \lambda h_{\alpha
  \beta} + O(\lambda^2) \,\, ,
\end{equation}
where
\begin{equation}
\label{eq:perturbations-series} 
h_{\alpha \beta} = \frac{c_{\alpha \beta}(t,\theta,\phi)}{r} + O(1) \,\, .
\end{equation}
where in eq.(\ref{eq:perturbations-series}), the term ``$O(1)$'' denotes
a term that, when multiplied by $r^\epsilon$ for any $\epsilon > 0$, vanishes
as $r \rightarrow 0$. 

Now, by assumption (i) of section \ref{sec:example}, for each $\lambda$,
$g_{ab} (\lambda)$ is a vacuum solution of Einstein's equation for
$r > \lambda \bar{R}$ and is jointly smooth in $(\lambda, x^\alpha)$
in this coordinate range. It follows that for all $r > 0$,
$h_{ab}$ is a solution the linearized Einstein equation off of 
$g_{ab} (\lambda=0)$, i.e.,
\begin{equation}
G^{(1)}_{ab}[h_{cd}] = -\frac{1}{2} \nabla_a \nabla_b h^c_{\ c} - \frac{1}{2}\nabla^c\nabla_c h_{ab} + \nabla^c \nabla_{(b}h_{a)c} = 0, \,\,  r>0 \,\, ,
\label{lee}
\end{equation}
where here and in the following, $\nabla_a$ denotes the derivative
operator associated with $g_{ab} (\lambda=0)$, and indices are raised
and lowered with $g_{ab} (\lambda=0)$. Equation (\ref{lee}) holds only
for $r > 0$, and, indeed, if $c_{\alpha \beta} \neq 0$, $h_{ab}$ is
singular at $r = 0$. However, even if $c_{\alpha \beta} \neq 0$, the
singularity of each component of $h_{ab}$ is locally integrable with
respect to the volume element associated with $g_{ab}(\lambda=0)$,
i.e., each component, $h_{\alpha \beta}$, is a locally $L^1$ function
on the entire spacetime manifold, including $r=0$.  {\it Thus,
  $h_{ab}$ is well defined as a distribution on all of spacetime.}
The quantity $T_{ab} \equiv G^{(1)}_{ab}[h_{cd}]/8 \pi$ is therefore
automatically well defined as a distribution. This quantity has the
interpretation of being the ``source'' for the metric perturbation
(\ref{eq:perturbations-series})---even though all of our spacetimes
$g_{ab}(\lambda)$ for $\lambda > 0$ have excluded the ``source region'' $r
\leq \lambda \bar{R}$. It follows immediately from eq.(\ref{lee})
that, as a distribution, $T_{ab}$ must have support on $\gamma$ in the
sense that it must vanish when acting on any test tensor field
$f^{ab}$ whose support does not intersect $\gamma$. We now compute
$T_{ab}$.

By definition, $T_{ab} \equiv G^{(1)}_{ab}[h_{cd}]/8 \pi$ is the distribution
on spacetime whose action on an arbitrary smooth,
compact support, symmetric tensor field $f_{cd} = f_{dc}$ is given by 
\begin{equation}
\label{eq:distributional-equation}
8 \pi T(f) = \int_M G^{(1)}_{ab} [f_{cd}] h^{ab} \sqrt{-g} d^4x = 0 \,\, ,
\end{equation}
where $\sqrt{-g} d^4x$ denotes the volume element associated with
$g_{ab}(\lambda=0)$ and we have used the fact that the operator
$G^{(1)}_{ab}$ is self-adjoint\footnote{See \cite{wald-prl} for the
  definition of adjoint being used here. If $G^{(1)}_{ab}$ were not
  self-adjoint, then the adjoint of $G^{(1)}_{ab}$ would have appeared
  in eq.~(\ref{eq:distributional-equation}).}. Note that the right
side of this equation is well defined since $G^{(1)}_{ab} [f_{cd}]$ is
a smooth tensor field of compact support and $h^{ab}$ is locally
$L^1$. We can evaluate the right side of
eq.~(\ref{eq:distributional-equation}) by integrating over the region
$r > \epsilon > 0$ and then taking the limit as $\epsilon \rightarrow
0$.  In the region $r > \epsilon$, $h^{ab}$ is
smooth, and a straightforward ``integration by parts'' calculation
shows that
\begin{equation}
\label{eq:adjoint-relation}
G^{(1)}_{ab} [f_{cd}] h^{ab} - G^{(1)}_{ab} [h_{cd}] f^{ab} = \nabla_c s^c \,\, ,
\end{equation}
where
\begin{multline}
\label{eq:adjoint-current} 
s^c = h^{ab} \nabla^c f_{ab} - \nabla^c h^{ab} f_{ab} + h^{bc}\nabla_b f - \nabla_b h^{bc} + 2 \nabla^a h^{bc}f_{ab} - 2 h_{ab} \nabla^a f^{bc} \\ + \nabla^c h f - h \nabla^c f + h \nabla_b f^{bc} - \nabla_b h f^{bc}\,\, ,
\end{multline}
where $f = f_{ab} g^{ab}(\lambda = 0)$. 
Since $G^{(1)}_{ab} [h_{cd}] = 0$ for $r > 0$, it follows immediately that
\begin{equation}\label{eq:distributional-equation2}
T(f) = \frac{1}{8 \pi} \lim_{\epsilon \rightarrow 0} \int_{r > \epsilon} G^{(1)}_{ab} [f_{cd}] h^{ab} = \frac{1}{8 \pi} \lim_{\epsilon \rightarrow 0} \int_{r=\epsilon} s^a n_a dS \,\, .
\end{equation}
Using eqs.~(\ref{eq:perturbations-series}) and (\ref{eq:adjoint-current}),
we find that $T(f)$ takes the form
\begin{equation}
T(f) = \int dt N_{ab}(t) f^{ab}(t,r=0)\,\, ,
\label{TN}
\end{equation}
where $N_{ab}(t)$ is a smooth, symmetric ($N_{ab} = N_{ba}$) 
tensor field on $\gamma$ whose
components are given in terms of suitable angular averages of
$c_{\alpha \beta}$ and its first angular derivatives. In other words,
the distribution $T_{ab}$ is given by\footnote{In fact, by our
  coordinate choice, we have $\sqrt{-g} = 1$ on $\gamma$ and $\frac{d\tau}{dt}=1$, but we prefer to leave in these factors so that this formula holds for an arbitrary choice of coordinates.}
\begin{equation}
T_{ab} = N_{ab}(t) \frac{\delta^{(3)}(x^i)}{\sqrt{-g}} \frac{d\tau}{dt}\,\, ,
\label{eq:delta-source}
\end{equation}
where $\delta^{(3)}(x^i)$ is the ``coordinate delta-function'' (i.e.,
$\int \delta^{(3)}(x^i) d^3 x^i = 1$).

We now use the fact that, since the differential operator 
$G^{(1)}_{ab}$ satisfies the linearized
Bianchi identity $\nabla^a G^{(1)}_{ab} = 0$, the distribution
$T_{ab}$ must satisfy $\nabla^a T_{ab} = 0$ in the distributional sense. 
This means that the
action of $T_{ab}$ must vanish on any test tensor field of the form
$f_{ab} = \nabla_{(a} f_{b)}$ where $f_a$ is smooth and of compact
support. In other words, by eq.~(\ref{TN}), the tensor field $N_{ab}$
on $\gamma$ must be such that for an arbitrary smooth vector field
$f^a$ on spacetime, we have
\begin{equation}
\int dt N_{ab}(t) \nabla^a f^b(t,r=0) = 0 \,\, .
\label{Nf}
\end{equation}
Now for any $i=1,2,3$, choose $f^a$ to have components of the form
$f^\mu = x^i F(x^1, x^2, x^3) c^\mu(t)$, where each $c^\mu$ ($\mu = 0,1,2,3$)
is an arbitrary
smooth function of compact support in $t$ and $F$ is a smooth
function of compact spatial support, with $F=1$ in a neighborhood of
$\gamma$. Then eq.~(\ref{Nf}) yields
\begin{equation}
\int dt N_{i\mu}(t) c^\mu(t) = 0 
\end{equation}
for all $c^\mu(t)$, which immediately implies that $N_{i \mu} = N_{\mu
  i} = 0$ for all $i=1,2,3$ and all $\mu = 0,1,2,3$. In other words, we have
shown that $N_{ab}(t)$ must take the form
\begin{equation}
N_{ab} = M(t) u_a u_b
\label{N}
\end{equation}
where $u^a$ denotes the unit tangent to $\gamma$, i.e., $u^a$ is the 4-velocity
of $\gamma$. Now choose $f^a$ to be an arbitrary
smooth vector field of compact support. Then eqs.~(\ref{Nf}) and (\ref{N})
yield
\begin{equation}
0 = \int dt M(t)u_b (u_a \nabla^af^b) = - \int dt u^a \nabla_a(M(t)
u_b) f^b \,\, ,
\end{equation}
where we integrated by parts in $t$ to obtain the last equality. Since
$f^a$ is arbitrary, this immediately implies that
\begin{equation}
u^a \nabla_a(M(t) u_b) = 0 \,\, .
\end{equation}
This, in turn, implies that
\begin{equation}
dM/dt = 0 \,\, ,
\end{equation}
i.e., $M$ is a constant along $\gamma$, and, if $M \neq 0$,
\begin{equation}
u^a \nabla_a u^b = 0 \,\, ,
\end{equation}
i.e., in the case where $M \neq 0$, $\gamma$ is a geodesic of
$g_{ab}(\lambda = 0)$, as we desired to show \footnote{Some previous
  derivations \cite{d'eath,mino-sasaki-tanaka,poisson} of geodesic
  motion do not appear to make explicit use of the fact that $M \neq
  0$. It is critical that this assumption be used in any valid
  derivation of geodesic motion, since a derivation that
  holds for $M=0$ effectively would show that all curves are
  geodesics.}.

In summary, we have shown that for any one-parameter family of metrics
$g_{ab}(\lambda)$ satisfying assumptions (i)-(iii) of section 
\ref{sec:example}, to first order in $\lambda$, the far zone metric
perturbation $h_{ab}$ corresponds to a solution to the linearized Einstein
equation with a point particle source
\begin{equation}
T_{ab} = M u_a u_b \frac{\delta^{(3)}(x^i)}{\sqrt{-g}} \frac{d\tau}{dt}\,\, ,
\label{ptparticle}
\end{equation}
where $M$ is a constant and $u^a$ is the 4-velocity of $\gamma$, which
must be a geodesic if $M \neq 0$. We refer to $M$ as the {\it mass} of
the particle.
It is rather remarkable that the
point particle source (\ref{ptparticle}) is an {\it output} of our
analysis rather than an input. Indeed, we maintain that the result 
we have just derived is
what provides the justification for the notion of ``point
particles''---a notion that has played a dominant role in classical
physics for more than three centuries. In fact, the notion of point
particles makes no mathematical sense in the context of nonlinear field
theories like general relativity. Nevertheless, we have just shown
that the notion of a (structureless) ``point particle'' arises
naturally as an {\it approximate} description of sufficiently small
bodies---namely, a description that is valid to first order in
$\lambda$ in the far zone for arbitrary one-parameter families
of metrics $g_{ab}(\lambda)$ satisfying the assumptions of
section \ref{sec:example}. This description is valid independently of
the nature of the ``body'', e.g., it holds with equal validity 
for a small black hole as for a small blob of ordinary matter.

\section{Description of Motion to First Order in $\lambda$} \label{sec:dipole}

In the previous section, we established that, to zeroth order in
$\lambda$, any body (or black hole) of nonvanishing mass moves on a
geodesic of the background spacetime. Much of the remainder of this
paper will be devoted to finding the corrections to this motion, valid
to first order in $\lambda$ in the far zone. In this section, we
address the issue of what is meant by the ``motion of the body'' to
first order in $\lambda$.

The first point that should be clearly recognized is that it is far
from obvious how to describe ``motion'' in terms of a worldline for
$\lambda > 0$. Indeed, the metric $g_{ab}(\lambda)$ is defined only
for $r > \lambda \bar{R}$, so at finite $\lambda$ the spacetime
metric may not even be defined in a neighborhood of $\gamma$. If we
were to assume that $\bar{R} >> M$ and that the region $r < \lambda
\bar{R}$ were ``filled in'' with sufficiently ``weak field
matter''---so that $\bar{\mathcal R} \bar{R}^2 << 1$, where
$\bar{\mathcal R}$ denotes the supremum of the components of the
Riemann curvature tensor of $\bar{g}_{ab}(\lambda)$ in the ``filled
in'' region---then it should be possible to define a ``center of
mass'' worldline at finite $\lambda$, and we could use this worldline
to characterize the motion of the body \cite{beiglbock}.  However, we do not
wish to make any ``weak field'' assumptions here, since we wish to
describe to motion of small black holes and other ``strong field''
objects. Since it is not clear how to associate a worldline to the
body at finite $\lambda$---and, indeed, the ``body'' is excluded from
the spacetime region we consider at finite $\lambda$---it is not clear
what one would mean by a ``perturbative correction'' to $\gamma$ to
first or higher order in $\lambda$.

A second point that should be understood is that if we have succeeded in
defining the worldlines describing the motion the body at finite $\lambda$,
\begin{equation}
x^i(\lambda, t) = z^i(\lambda,t) = \lambda Z^i(t) + O(\lambda^2) \,\, ,
\end{equation}
then the ``first order in $\lambda$ perturbative correction'', $Z^i$,
to the zeroth order motion $\gamma$ (given by $x^i(t) = 0$) is most
properly viewed as the spatial components of a vector field, $Z^a$,
defined along $\gamma$. This vector field describes the
``infinitesimal displacement'' to the corrected motion to first order
in $\lambda$. The time component, $Z^0$, of $Z^a$ depends on on how we
identify the time parameter of the worldlines at different values of
$\lambda$ and is not physically relevant; we will set $Z^0 = 0$ so
that $Z^a$ is orthogonal to the tangent, $u^a$, to $\gamma$ in the
background metric $g_{ab}(\lambda = 0)$. Thus, when we seek equations
of motion describing the first order perturbative corrections to
geodesic motion, we are seeking equations satisfied by the vector field
$Z^a(t)$ on $\gamma$.

A third point that should be clearly recognized is that $Z^a$ and any equations
of motion satisfied by $Z^a$ will depend on our
choice of gauge for $h_{ab}$. To
see this explicitly, suppose that
we perform a smooth\footnote{Gauge transformations where $A^\mu$ is
  not smooth at $x^i = 0$ are also permitted under the coordinate
  freedom stated at the end of section \ref{sec:example}. However, it
  suffices to consider smooth $A^\nu$ for our consderations here. 
The change in the 
description of motion under non-smooth gauge transformations 
will be treated in Appendix A.} gauge
transformation of the form
\begin{equation}
\label{eq:transform}
x^\mu \rightarrow 
\hat{x}^\mu = x^\mu - \lambda A^\mu(x^\nu) + O(\lambda^2) \,\, .
\end{equation}
Under this gauge transformation, we have
\begin{equation}
h_{\mu \nu} \rightarrow \hat{h}_{\mu \nu} = h_{\mu \nu} + 2\nabla_{(\mu} A_{\nu)} \,\, .
\end{equation}
However, clearly, the new description of motion will be of the form \cite{barack-ori}
\begin{equation}
\hat{x}^i(\lambda,t) = \hat{z}^i (\lambda, t)
\end{equation}
where
\begin{equation}
\hat{z}^i(t) = z^i(t) - \lambda A^i(t,x^j=0) +O(\lambda^2) \,\, .
\end{equation}
Thus, we see that $Z^a$ transforms as
\begin{equation}
Z^i(t) \rightarrow \hat{Z}^i(t) = Z^i(t) - A^i(t,x^j=0)
\label{smoothgauge}
\end{equation}
in order that it describe the same perturbed motion. Thus, the first
order correction, $Z^a(t)$, to the background geodesic motion contains
no meaningful information by itself and, indeed, it can always be set
to zero by a smooth gauge transformation. Only the pair $(h_{ab}, Z^a(t))$
has gauge invariant meaning.

We turn now to the definition of the first order perturbed motion. Our
definition relies on the fact, proven in section \ref{sec:assumptions}
above, that for each $t_0$, $\bar{g}_{\bar{\mu}
  \bar{\nu}}(\lambda=0;t_0;\bar{x}^\alpha)$ is a stationary,
asymptotically flat spacetime. Therefore, $\bar{g}_{ab}(\lambda=0)$
has well defined sets of mass (``electric parity'') and angular
momentum (``magnetic parity'') multipole moments \cite{geroch-multipoles, hansen} and, indeed, the spacetime is characterized by the values of these two sets of multipole moments \cite{beig-simon,kundu}. The multipole moments (other than the lowest nonvanishing multipoles of each type) depend upon a choice of conformal factor \cite{geroch-multipoles, hansen}, which, rougly speaking,corresponds to a choice of ``origin''.  We choose the conformal factor $\Omega = 1/\bar{r}^2$ to define all of the multipoles, corresponding
to choosing the origin at $\bar{r} = 0$.  For a metric of the form
eq.~(\ref{eq:fullnearzone2})---with $(a_{\mu \nu})_{00} = \eta_{\mu
  \nu}$ by our coordinate choice imposed in the previous section that
$g_{\mu \nu}(\lambda = 0) = \eta_{\mu \nu}$ on $\gamma$---the mass
will be simply the $l=0$ part of the coefficient of $1/\bar{r}$ in the
large $\bar{r}$ expansion of $\frac{1}{2} \bar{g}_{\bar{t} \,
  \bar{t}}(\lambda=0;t_0)$. Similarly, the mass dipole moment will be the coefficient of the $l=1$ part of this quantity at order $1/\bar{r}^2$.

It is well known that if the mass is nonzero, the mass dipole moment
is ``pure gauge'' and can be set to zero by choice of conformal
factor/``origin''. We now explicitly show that, with our choice of conformal
factor $\Omega = 1/\bar{r}^2$, the mass dipole moment can be set to
zero by a smooth gauge transformation of the form
(\ref{eq:transform}). It follows from the linearized Einstein equation
(\ref{lee}) with source (\ref{ptparticle}) applied to $h_{ab}$,
eq.~(\ref{eq:perturbations-series}), that the time-time component of
$h_{ab}$ takes the form 
\begin{equation}
h_{tt} = \frac{2M}{r} + O(1) \,\, ,
\end{equation}
i.e., in the notation of eq.~(\ref{eq:rseries}), we have
$(a_{tt})_{01} = 2M$.  Comparing with eq.~(\ref{eq:fullnearzone}) (and also
using the fact that $(a_{tt})_{00} = -1$), we see that at each $t_0$
\begin{equation}
\bar{g}_{\bar{t} \, \bar{t}}(\lambda=0;t_0) = - (1 - \frac{2M}{\bar{r}}) +
O(1/\bar{r}^2) \,\, .
\end{equation}
From this equation, we see that the ``particle mass'', $M$, of the
``source'' of the far zone metric perturbation (see
eq.~(\ref{ptparticle})) is also the Komar/ADM mass of the stationary,
asymptotically flat spacetime $\bar{g}_{ab}(\lambda=0;t_0)$.  We now
calculate the effect of the coordinate transformation
(\ref{eq:transform}) on $\bar{g}_{\bar{t}
  \bar{t}}(\lambda=0;t_0)$. The transformation (\ref{eq:transform})
corresponds to changing the barred coordinates by
\begin{equation}
\label{eq:transformbar}
\bar{x}^\mu \rightarrow
\hat{\bar{x}}^\mu = \bar{x}^\mu - A^\mu(t_0, x^i=0) + O(\lambda) \,\, ,
\end{equation}
i.e., to zeroth order in $\lambda$, it corresponds to a ``constant
displacement'' of coordinates. Since
\begin{equation}\label{eq:displacement}
\frac{1}{\bar{r}} = \frac{1}{|\hat{\bar{x}}^i + A^i(t_0,0)|} =
\frac{1}{\hat{\bar{r}}} - \frac{A_i \hat{\bar{x}}^i}{\hat{\bar{r}}^3}
+O(1/\hat{\bar{r}}^3) \,\, ,
\end{equation}
it can be seen that this transformation have the effect of changing
the mass dipole moment by $-M A^i$. In particular,
we can always choose $A^i$ so as to set the mass dipole moment to zero.

Now, the ``near zone'' coordinates $\hat{\bar{x}}^i$ for which the
mass dipole moment vanishes have the interpretation of being ``body
centered'' coordinates to zeroth order in $\lambda$. The origin
$\hat{x}^i = 0$ of the corresponding ``far zone'' coordinates
$\hat{x}^i$ therefore has the interpretation of representing the
``position'' of the center of mass of the body to first order in
$\lambda$. We shall use this to define the correction to geodesic
motion to first order in $\lambda$ by proceeding as follows:

First, we shall choose our coordinates, $x^\mu$, to zeroth order in
$\lambda$ by choosing convenient coordinates for the ``background
spacetime'' $g_{ab}(\lambda = 0)$. (We will use Fermi normal
coordinates based on $\gamma$.) Next, we will define our coordinates,
$x^\mu$, to first order in $\lambda$ by choosing a convenient gauge
for $h_{ab}$, eq.~(\ref{eq:perturbations-series}). (We will choose the
Lorenz gauge $\nabla^a (h_{ab} - \frac{1}{2} h g_{ab}) = 0$.) Then we will
introduce the smooth coordinate transformation (\ref{eq:transform}), and
impose the requirement that $A^\mu$ be such that the mass dipole moment
of $\bar{g}_{ab}(\lambda=0;t_0)$ vanish for all $t_0$. Since the ``location''
of the body in the new coordinates is $\hat{z}^i(t) = 0$, the 
first order perturbative correction $Z^a(t)$ to the
motion of the
body in our original coordinates $x^\mu$ will be given by
\begin{equation}
Z^i(t) = A^i(t,x^j=0)
\end{equation}
Of course, the particular $Z^a(t)$ that we obtain in any given case
will depend upon the particular one-parameter family $g_{ab}(\lambda)$
that we consider. What is of interest is any ``universal relations''
satisfied by $Z^a(t)$ that are independent of the choice of
one-parameter family satisfying assumptions (i)-(iii) of section
\ref{sec:example}. Such universal relations would provide us with
``laws of motion'' for point particles that take self-force effects
into account. In the next section, we will show (via a lengthy
calculation) that such a universal relation exists for $d^2Z^i/dt^2$,
thus providing us with general ``equations of motion'' for all ``point
particles'', valid to first order in $\lambda$.

Finally, we note that if we wish to describe motion beyond first order
in $\lambda$, it will be necessary to define a ``representative worldline'' in the far zone to at least second order in $\lambda$. We shall
not attempt to do so in this paper. The definition of a suitable
representative worldline is probably the greatest obstacle to
extending the results of this paper to higher order in perturbation
theory.

\section{Computation of Perturbed Motion} \label{sec:calculation}

In the section \ref{sec:geodesic} we found that first-order far zone
perturbations of the background spacetime $g_{ab}(\lambda = 0)$ are
sourced by a point particle stress-energy, eq.~(\ref{ptparticle}).
For the remainder of this paper, we will assume that $M \neq 0$, so
that, as shown in section \ref{sec:geodesic}, the lowest order
motion is described by a geodesic, $\gamma$, of the background
spacetime. We will need expressions for the components of the far zone
metric, $g_{\mu \nu}|_{\lambda = 0}$, its first order perturbation,
$h_{\mu \nu} \equiv \partial g_{\mu \nu}/ \partial \lambda |_{\lambda
  = 0}$, and its second order perturbation $j_{\mu \nu} \equiv
\frac{1}{2} \partial^2 g_{\mu \nu}/ \partial \lambda^2 |_{\lambda =
  0}$. It is convenient to choose our coordinates $x^\mu$
to zeroth order in $\lambda$
to be Fermi normal coordinates with respect to 
the background geodesic
$\gamma$, and choose these coordinates to first order
in $\lambda$ so that $h_{\mu \nu}$ satisfies the Lorenz gauge condition
$\nabla^b \tilde{h}_{ab} = 0$, where $\tilde{h}_{ab} \equiv 
h_{ab} - \frac{1}{2} h g_{ab}|_{\lambda = 0}$ with $h \equiv h_{ab}
g^{ab}|_{\lambda = 0}$. Then the linearized Einstein equation reads
\begin{eqnarray}
\label{eq:wave2}
\nabla^c \nabla_c \tilde{h}_{ab} - 2 R^c{}_{ab}{}^d \tilde{h}_{cd} &=& - 16 \pi M
\int_\gamma \delta_4(x,z(\tau))\,u_a(\tau)u_b(\tau)\,d\tau, \\
\nabla^b \tilde{h}_{ab} &=& 0.
\label{eq:gauge2}
\end{eqnarray}

This system of equations can be solved using the Hadamard expansion
techniques of DeWitt and Brehme
\cite{dewitt-brehme,mino-sasaki-tanaka,quinn-wald}.  Since this
technology has been used in all previous derivations of gravitational
self-force, we do not review it here but simply present results.
Equation (2.27) of Mino, Sasaki, and Tanaka \cite{mino-sasaki-tanaka}
provides a covariant expression for the perturbations in terms of
parallel propagators and Synge's world function on the background
metric (see, e.g., reference \cite{poisson} for definitions of these
quantities). The Fermi normal coordinate components of these
tensors are easily calculated with the aid of expressions from section
8 of Poisson \cite{poisson}.  Combining this with the form of the
background metric in Fermi normal coordinates, we obtain
\begin{align}\label{eq:metric}
g_{\alpha \beta}(\lambda;t,x^i) & = \eta_{\alpha \beta} + B_{\alpha i
  \beta j}(t)x^i x^j + O(r^3) \nonumber \\ & + \lambda \left(
\frac{2M}{r}\delta_{\alpha \beta} + h^{\textrm{\tiny
    tail}}_{\alpha \beta}(t,0) + h^{\textrm{\tiny tail}}_{\alpha
  \beta i}(t,0)x^i + M \mathcal{R}_{\alpha \beta}(t,x^i) + O(r^2) \right) +
O(\lambda^2) \,\, ,
\end{align}
where the quantities $B_{\alpha \beta \gamma \delta}$ and
$\mathcal{R}_{\alpha \beta}$ are defined by the following expressions
in terms of the Fermi normal
coordinate components of the Riemann tensor of the background metric
\begin{align}
B_{0 k 0 l} & = - R_{0 k 0 l} & \mathcal{R}_{0 0} & = 7 R_{0 k 0 l}
\frac{x^k x^l}{r}\\ B_{i k 0 l} & = -\frac{2}{3} R_{i k 0 l} &
\mathcal{R}_{i 0} & = \frac{2}{3} R_{i k 0 l} \frac{x^k x^l}{r} - 2
R_{i 0 k 0} x^k \\ B_{i k j l} & = -\frac{1}{3} R_{i k j l} &
\mathcal{R}_{i j} & = -\frac{13}{3} R_{i k j l} \frac{x^k x^l}{r} - 4
r R_{i 0 j 0},
\end{align}
and $h_{\alpha \beta}^{\textrm{\tiny tail}}$ and $h_{\alpha \beta \gamma}^{\textrm{\tiny tail}}$ are given by 
\begin{align}\label{eq:tail}
h_{\alpha \beta}^{\textrm{\tiny tail}}(x) & \equiv M \int_{-
  \infty}^{\tau_{\textrm{ret}}^-} \left( G_{+ \alpha \beta \alpha ' \beta '} -
\frac{1}{2} g_{\alpha \beta}G_{+ \ \gamma \alpha ' \beta '}^{\ \gamma
} \right) (x,z(\tau'))u^{\alpha '} u^{\beta '}d \tau ', \\
\label{eq:tail2}
h_{\alpha \beta \gamma}^{\textrm{\tiny tail}}(x) & \equiv M \int_{-
  \infty}^{\tau_{\textrm{ret}}^-} \nabla_\gamma \left( G_{+ \alpha \beta \alpha ' \beta '} -
\frac{1}{2} g_{\alpha \beta}G_{+ \ \delta \alpha ' \beta '}^{\ \delta
} \right)(x,z(\tau')) u^{\alpha '} u^{\beta '}d \tau '.
\end{align}
In these expressions, $G_+$ is the Lorenz gauge retarded Green's
function, normalized with a factor of $-16\pi$, following
\cite{quinn-wald}.  As previously mentioned, the symbol $\tau_{\textrm{ret}}^-$ indicates that the range of the integral extends just short of the retarded time $\tau_{\textrm{ret}}$, so that only the ``tail'' (i.e., interior of the light cone) portion of the Green's function is used (see, e.g., reference \cite{poisson} for details).  We define $h_{\alpha \beta \gamma}^{\textrm{\tiny tail}}$, rather than working with derivatives of $h_{\alpha \beta}^{\textrm{\tiny tail}}$, because $h_{\alpha \beta}^{\textrm{\tiny tail}}$ is not differentiable on the worldline.  (However, this non-differentiability is limited only to spatial derivatives of spatial components of $h_{\alpha \beta}^{\textrm{\tiny tail}}$\footnote{This can be seen from the fact that differentiation of $h_{\alpha \beta}^{\textrm{\tiny tail}}$ on the worldline $\gamma$ yields $h_{\alpha \beta \gamma}^{\textrm{\tiny tail}}$ plus the coincidence limit of the integrand of \eqref{eq:tail}, which is proportional to $R_{\alpha 0 \beta 0}$ times the gradient of $\tau_{\textrm{ret}}$.}, so that expressions expressions like \eqref{eq:MiSaTaQuWa-intro} are well-defined.)  A choice of retarded solution (corresponding to ``no
incoming radiation'') was made in writing these equations.  This choice
is not necessary, and one could add an arbitrary smooth solution
$h_{\alpha \beta}$ of the linearized Einstein equation to the first
order in $\lambda$ term on the right
side of eq.~(\ref{eq:metric}), which could then be carried through all
of our calculations straightforwardly. However, for simplicity, we will not
consider the addition of such a term.

Our derivation of gravitational self-force to first order in $\lambda$
will require consideration of second-order metric perturbations, so we
will have to carry the expansion of $g_{ab}(\lambda)$ somewhat beyond
eq.~(\ref{eq:metric}). (This should not be surprising in view of fact
that our above derivation in section \ref{sec:geodesic} of geodesic
motion at zeroth order in $\lambda$, required consideration of
first-order metric perturbations.)  In particular, we will need an explicit
expression for the quantity $(a_{\mu \nu})_{02}$ appearing in the far
zone expansion eq.~(\ref{eq:rseries}), i.e., the term of order
$\lambda^2$ that has the most singular behavior in $1/r$ (namely,
$1/r^2$).

The second-order perturbation $j_{ab}$ satisfies the second-order
Einstein equation, which takes the form
\begin{equation}\label{eq:e2}
G^{(1)}_{ab}[j] = - G^{(2)}_{ab}[h,h]
\end{equation}
where $G^{(2)}_{ab}$ denotes the second order Einstein tensor about the 
background metric $g^{ab}|_{\lambda = 0}$. Since the $O(1/r)$ part of $h_{ab}$
corresponds to the linearized Schwarzschild metric in isotropic coordinates
(see eq.~(\ref{eq:metric})),
it is clear that there is a particular solution to eq.~(\ref{eq:e2}) of the form
\begin{equation}
j^I_{\alpha \beta} = \frac{M^2}{2 r^2} \left( 3 \eta_{\mu \nu} - t_{\mu}t_{\nu} \right) + O(r^{-1}) 
\label{JI}
\end{equation}
as $r \rightarrow 0$, where 
\begin{equation}
n^i \equiv x^i/r
\label{ni}
\end{equation}
and $n^0 = 0$, whereas $t_\alpha \equiv \delta_{\alpha 0}$.
(The explicitly written term on the right side of eq.~(\ref{JI}) is
just the $O(M^2)$ part of the Schwarzschild metric in isotropic coordinates.)
The general solution to eq.~(\ref{eq:e2}) can then be written
as
\begin{equation}
j_{ab} = j^I_{ab} + j^H_{ab} \,\, ,
\end{equation}
where $j^H_{ab}$ is a homogeneous solution of the linearized Einstein
equation.  We wish to compute the $O(1/r^2)$ part of $j^H_{ab}$, i.e.,
writing
\begin{equation}
j^H_{ab} = \frac{C_{ab}(t,\theta,\phi)}{r^2} + O(r^{-1}) \,\, ,
\label{jH}
\end{equation}
we wish to compute $C_{ab}$. Now, although the equations of motion to
first order in $\lambda$ depend upon a choice of gauge to first order in 
$\lambda$ (see section \ref{sec:dipole}), they cannot depend upon a choice
of gauge to second 
order in $\lambda$, since a second order gauge transformation cannot
affect the mass dipole moment of the background scaled metric
$\bar{g}_{\bar{\mu} \bar{\nu}} (\lambda=0)$. [Note added: Please see arXiv:1104.5205.] (We have also verified by 
a direct, lengthy computation
that second order gauge transformations
do not produce changes in the equations of motion to first order in $\lambda$.)
Therefore, we are free to
impose any (admissible) second order gauge condition on $j^H_{ab}$.
It will be convenient to require that the Lorenz gauge condition
$\nabla^a (j^H_{ab} - \frac{1}{2} j^H g_{ab}) = 0$ be satisfied to
order $1/r^3$. The $O(1/r^4)$ part of the linearized Einstein equation
together with the $O(1/r^3)$ part of the Lorenz gauge condition then
yields
\begin{align}
\partial^i \partial_i \left( \frac{1}{r^2} \tilde{C}_{\mu \nu}(t,\theta,\phi) \right) & = 0 \\
\partial^i \left( \frac{1}{r^2} \tilde{C}_{i\mu}(t,\theta,\phi) \right) & = 0.
\end{align}
where $\tilde{C}_{ab} = C_{ab} - \frac{1}{2} C \eta_{ab}$. This system
of equations for $\frac{1}{r^2} C_{\mu \nu}$ is the same system of
equations as is satisfied by stationary solutions of the flat spacetime
linearized Einstein equation (but our $C_{\mu \nu}$ may depend on
time). The general solution of these equations is $\tilde{C}_{ij} =
0$, $\tilde{C}_{i0} = F(t)n_i + 2 S_{ij}(t)n^j$, and $\tilde{C}_{00} =
4P_i(t)n^i$, where $S_{ij}$ is antisymmetric, $S_{ij} = - S_{ji}$,
where $F$, $S_{ij}$, and $P_i$ have no spatial dependence, and where $n^i$
was defined by eq.~(\ref{ni}). By a further second order gauge
transformation (of the form $\xi_\mu = \delta_{\mu 0} F(t)/r $), we can
set $F(t) = 0$. We thus obtain
\begin{align}
C_{00}(t,\theta,\phi) &= 2 P_i(t) n^i(\theta,\phi) \\
C_{i0}(t,\theta,\phi) &= 2 S_{ij}(t) n^j(\theta,\phi) \\
C_{ij}(t,\theta,\phi) &= 2 \delta_{ij} P_k(t) n^k(\theta,\phi) \,\, ,
\end{align}
which is of the same form as the general stationary $l=1$ perturbation
of Minkowski spacetime (see, e.g., \cite{zhang}), except that time
dependence is allowed for $S_{ij}$ and $P_i$. As we shall see shortly,
$S_{ij}$ and $P_i$ correspond, respectively, to the spin and mass
dipole moment of the body.

We now may write for the metric through $O(\lambda^2)$,
\begin{equation}\label{eq:metric2}
\begin{split}
g_{\alpha \beta}(\lambda;t,x^i) & =  \eta_{\alpha \beta} + B_{\alpha i
  \beta j}(t)x^i x^j + O(r^3) \\ & \quad +  \lambda \left(
\frac{2M}{r}\delta_{\alpha \beta} + h^{\textrm{\tiny
    tail}}_{\alpha \beta}(t,0) + h^{\textrm{\tiny tail}}_{\alpha
  \beta i}(t,0)x^i + M \mathcal{R}_{\alpha \beta}(t) + O(r^2) \right) 
\\ & \quad +  \lambda^2 \biggl( \frac{M^2}{2 r^2} \left( 3 \eta_{\mu \nu} - t_{\mu}t_{\nu} \right) + \frac{2}{r^2} P_i(t) n^i \delta_{\alpha
  \beta} - \frac{4}{r^2} t_{(\alpha} S_{\beta) j}(t) n^j \\
& \qquad \qquad + \frac{1}{r}
K_{\alpha \beta}(t,\theta,\phi) + H_{\alpha \beta}(t,\theta,\phi) +
O(r) \biggr) + O(\lambda^3) 
\end{split}
\end{equation}
where we have introduced the unknown tensors $K$ and $H$, and
$S_{\alpha \beta}$ is the antisymmetric tensor whose spatial components
are $S_{ij}$ and whose time components vanish, i.e.,
\begin{equation}
S_{0i} = 0 \,\, .
\end{equation}

We now follow the strategy outlined in section \ref{sec:dipole}. We consider
a {\it smooth} coordinate shift of the form \eqref{eq:transform},
\begin{equation}\label{eq:transform2}
\hat{x}^\mu = x^\mu - \lambda A^\mu(x^\nu) + O(\lambda^2),
\end{equation}
and choose $A^\mu$ so as to make the mass dipole moment of
$\bar{g}_{\hat{\bar{\alpha}} \hat{\bar{\beta}}}(\lambda, t_0)$ vanish
for all $t_0$. A straightforward application of the coordinate
transformation (\ref{eq:transform2}) to the metric (\ref{eq:metric2})
yields
\begin{equation}\label{eq:metric-hat}
\begin{split}
g_{\hat{\alpha} \hat{\beta}} & = \eta_{\alpha \beta} + B_{\alpha i
  \beta j}(\hat{t})\hat{x}^i \hat{x}^j + O(r^3) \\ & + \lambda \left(
\frac{2 M}{\hat{r}}\delta_{\alpha \beta} + h^{\textrm{\tiny
    tail}}_{\alpha \beta}(\hat{t},0) + h^{\textrm{\tiny tail}}_{\alpha
  \beta i}(\hat{t},0)\hat{x}^i + M \mathcal{R}_{\alpha
  \beta}(\hat{t},\hat{x}^i) + 2 A_{(\alpha , \beta)}(\hat{t},\hat{x}^i) + 2
B_{\alpha i \beta j}(\hat{t})\hat{x}^i A^j(\hat{t},\hat{x}^i) + O(r^2)
\right) \\ & + \lambda^2 \biggl( \frac{M^2}{2 \hat{r}^2} \left(3 \eta_{\mu \nu} - t_{\mu}t_{\nu}\right) + \frac{2}{\hat{r}^2}
\left[P_i(\hat{t})-M A_i(\hat{t},0)\right] n^i \delta_{\alpha \beta} -
\frac{4}{\hat{r}^2} t_{(\alpha} S_{\beta) j}(\hat{t}) n^j \\ & \qquad \quad
+ \frac{1}{\hat{r}} K_{\alpha \beta}(\hat{t},\theta,\phi) + H_{\alpha
  \beta}(\hat{t},\theta,\phi) + O(r) \biggr) + O(\lambda^3),
\end{split}
\end{equation}
where we have ``absorbed'' the effects of the gauge transformation at
orders $\lambda^2r^{-1}$ and $\lambda^2r^0$ into the unknown tensors
$H,K$.  The corresponding ``near zone expansion'' (see
eqs.~(\ref{eq:rseries}), (\ref{bnmp}), and (\ref{eq:fullnearzone})) is
\begin{equation}\label{eq:scaled-metric}
\begin{split}
\bar{g}_{\hat{\bar{\alpha}} \hat{\bar{\beta}}}(\hat{t}_0) & =
\eta_{\alpha \beta} + \frac{2M}{\hat{\bar{r}}} \delta_{\alpha \beta} +
\frac{M^2}{2\hat{\bar{r}}^2} \left( 3 \eta_{\mu \nu} - t_{\mu}t_{\nu} \right) - \frac{4}{\hat{\bar{r}}^2} t_{(\alpha} S_{\beta) j}
n^j + \frac{2}{\hat{\bar{r}}^2} \left[P_i-M A_i\right] n^i
\delta_{\alpha \beta} + O \left( \frac{1}{\hat{\bar{r}}^3} \right)
\\ & \quad + \lambda \left[ h^{\textrm{\tiny tail}}_{\alpha \beta} + 2
  A_{(\alpha , \beta)} + \frac{1}{\hat{\bar{r}}} K_{\alpha \beta} +
  \frac{\hat{\bar{t}}}{\hat{\bar{r}}^2} \left( -4 t_{(\alpha} S_{\beta)
    j,0} n^j + 2 \left[P_{i,0}-M A_{i,0}\right] n^i \delta_{\alpha
    \beta} \right) + O \left( \frac{1}{\hat{\bar{r}}^2} \right) +
  \hat{\bar{t}} O\left( \frac{1}{\hat{\bar{r}}^3} \right) \right] \\ &
\quad + \lambda^2 \Biggl[ B_{\alpha i \beta j} \hat{\bar{x}}^i
  \hat{\bar{x}}^j + h^{\textrm{\tiny tail}}_{\alpha \beta
    \gamma}\hat{\bar{x}}^\gamma + M \mathcal{R}_{\alpha
    \beta}(\hat{\bar{x}}^i) + 2 B_{\alpha i \beta j}A^i
  \hat{\bar{x}}^j + 2 A_{(\alpha , \beta) \gamma} \hat{\bar{x}}^\gamma
  \\ & \qquad \qquad + H_{\alpha \beta} +
  \frac{\hat{\bar{t}}}{\hat{\bar{r}}} K_{\alpha \beta,0} +
  \frac{\hat{\bar{t}}^2}{\hat{\bar{r}}^2} \left( -4 t_{(\alpha} S_{\beta)
    j,00} n^j + 2 \left[P_{i,00}-M A_{i,00}\right] n^i \delta_{\alpha
    \beta} \right) \\ & \qquad \qquad + O\left(
  \frac{1}{\hat{\bar{r}}} \right) + \hat{\bar{t}} \ O \left(
  \frac{1}{\hat{\bar{r}}^3} \right) + \hat{\bar{t}}^2 \ O \left(
  \frac{1}{\hat{\bar{r}}^3} \right) \Biggr] + O(\lambda^3) \,\, .
\end{split}
\end{equation}
Notice that the indices on the left side of this equation have both a
``hat'' and ``bar'' on them to denote that they are components of
$\bar{g}_{ab}$ in the scaled coordinates associated with our new
coordinates $\hat{x}^\mu$. By contrast, the indices on the right side
have neither a ``hat'' nor a ``bar'', since they denote the
corresponding components in the unscaled, original coordinates
$x^\mu$. Thus, for example, $A_{\alpha , \beta}$ denotes the matrix of
first partial derivatives of the $x^\mu$-components of $A_a$ with
respect to the $x^\mu$ coordinates\footnote{Notice that the term
  $A_{(\alpha , \beta) \gamma}$ arises from Taylor expanding
  $A_{(\alpha , \beta)}$ with respect to the $\hat{x}^\mu$
  coordinates, so, in principle, the second partial derivative in this
  expression should be with respect to $\hat{x}^\gamma$ rather than
  $x^\gamma$.  However, since $\hat{x}^\gamma$ coincides with
  $x^\gamma$ at zeroth order in $\lambda$ and the $A_{(\alpha , \beta)
    \gamma}$ appears at second order in $\lambda$, we may replace the
  partial derivative with respect to $\hat{x}^\gamma$ by the partial
  derivative with respect to $x^\gamma$.}.  It also should be
understood that all tensor components appearing on the right side of
eq.~(\ref{eq:scaled-metric}) are evaluated at time $\hat{t}_0$,
and that $A_\alpha$ and its derivatives, as well as $h^{\textrm{\tiny tail}}_{\alpha \beta}$ and $h^{\textrm{\tiny tail}}_{\alpha \beta \gamma}$, are evaluated at
$\hat{x}^i=0$ (i.e., on the worldline $\gamma$).
Finally, the ``reversals'' in the roles of
various terms in going from the far zone expansion of the metric
eq.~(\ref{eq:metric2}) to the near zone expansion
eq.~(\ref{eq:scaled-metric}) should be noted. For example, the spin
term $\frac{-4}{r^2} t_{(\alpha} S_{\beta) j} n^j$ originated as a
second order perturbation in the far zone, but it now appears as part
of the background scaled metric in the near zone expansion. By
contrast, the term $B_{\alpha i \beta j}x^i x^j$ originated as part of
the background metric in the far zone, but it now appears as a second
order perturbation in the near zone expansion.

It is easy to see from eq.~(\ref{eq:scaled-metric})
that $P^i-MA^i$ is the mass dipole moment of
$\bar{g}_{\hat{\bar{\alpha}} \hat{\bar{\beta}}}$ at time $t_0$. We therefore set
\begin{equation}
A^i(t) = P^i(t)/M
\label{AP}
\end{equation}
for all $t$. Consequently,
no mass dipole term will appear in our expressions below.

Although we have ``solved'' for $A^i$ in eq.~(\ref{AP}), we have not
learned anything useful about the motion.\footnote{However, equation
  \eqref{AP} indicates clearly that solving for the displacement to
  center-of-mass coordinates $A^i$ is equivalent to simply determining
  the mass dipole moment $P^i$ in the original coordinates.  The computations of this section may
  therefore be recast as simply solving enough of the second-order
  perturbation equations for the mass dipole moment---and hence the
  motion---to be determined.} All useful information about $A^i$
will come from demanding that the metrics $g_{ab}(\lambda)$---or,
equivalently, $\bar{g}_{ab}(\lambda)$---be solutions of Einstein's
equation.  We may apply Einstein's equation perturbatively either via
the far zone expansion or the near zone expansion. The resulting
systems of equations are entirely equivalent, but the terms are
organized very differently. We find it more convenient to work with
the near zone expansion, and will do so below. We emphasize, however,
that we could equally well have used the far zone perturbation
expansion.  We also emphasize that no new information whatsoever can
be generated by matching the near and far zone expansions, since these
expansions have already been fully ``matched'' via eqs.
(\ref{eq:rseries}), (\ref{bnmp}), and (\ref{eq:fullnearzone}).

In the following, in order to make the notation less cumbersome, we
will drop the ``hat'' on the near-zone coordinates $\hat{\bar{x}}^\mu$
and on the components $\bar{g}_{\hat{\bar{\alpha}}
  \hat{\bar{\beta}}}$. No confusion should arise from this, since we
will not have occassion to use the original scaled coordinates
$\bar{x}^\mu$ below. On the other hand, we will maintain the ``hat''
on the coordinates $\hat{x}^\mu$, since we will have occassion to use
both $\hat{x}^\mu$ and $x^\mu$ below. Using this notation and setting
the mass dipole terms to zero, eq.~(\ref{eq:scaled-metric}) becomes
\begin{equation}\label{eq:scaled-metric2}
\begin{split}
\bar{g}_{\bar{\alpha} \bar{\beta}}(\hat{t}_0) & = \eta_{\alpha \beta}
+ \frac{2M}{\bar{r}} \delta_{\alpha \beta} + \frac{M^2}{2\bar{r}^2}
\left( 3 \eta_{\mu \nu} - t_{\mu}t_{\nu} \right) -
\frac{4}{\bar{r}^2} t_{(\alpha} S_{\beta) j} n^j + O \left(
\frac{1}{\bar{r}^3} \right) \\ & \quad + \lambda \left[
  h^{\textrm{\tiny tail}}_{\alpha \beta} + 2 A_{(\alpha , \beta)}
  + \frac{1}{r} K_{\alpha \beta} - 4 \frac{\bar{t}}{\bar{r}^2} t_{(\alpha} \dot{S}_{\beta) j} n^j + O \left( \frac{1}{\bar{r}^2}
  \right) + \bar{t} \ O\left( \frac{1}{\bar{r}^3} \right) \right] \\ &
\quad + \lambda^2 \biggl[ B_{\alpha i \beta j} \bar{x}^i \bar{x}^j +
  h^{\textrm{\tiny tail}}_{\alpha \beta \gamma}\bar{x}^\gamma +
  M \mathcal{R}_{\alpha \beta}(\bar{x}^i) + 2 B_{\alpha i \beta j}A^i
  \bar{x}^j + 2 A_{(\alpha , \beta) \gamma} \bar{x}^\gamma \\ & \qquad
  \qquad + H_{\alpha
    \beta} + \frac{\bar{t}}{\bar{r}} \dot{K}_{\alpha \beta} - 4 \frac{\bar{t}^2}{\bar{r}^2} t_{(\alpha} \ddot{S}_{\beta) j} n^j +O\left( \frac{1}{\bar{r}} \right) + \bar{t} \ O \left(
  \frac{1}{\bar{r}^3} \right) + \bar{t}^2 \ O \left(
  \frac{1}{\bar{r}^3} \right) \biggr] + O(\lambda^3) \,\, ,
\end{split}
\end{equation}
where the ``dots'' denote derivatives with respect to $t$.

We now apply the vacuum linearized Einstein equation---up to 
leading order, $1/\bar{r}^3$, in $1/\bar{r}$ as $\bar{r}
\rightarrow \infty$---to the first order term in $\lambda$ appearing
in eq.~(\ref{eq:scaled-metric2}), namely
\begin{equation}\label{eq:scaled-1}
\bar{g}^{(1)}_{\bar{\alpha} \bar{\beta}} = h^{\textrm{\tiny tail}}_{\alpha \beta}
+ 2 A_{(\alpha , \beta)} + \frac{1}{\bar{r}}K_{\alpha
  \beta}(\theta,\phi) - 4 
\frac{\bar{t}}{\bar{r}^2} t_{(\alpha} \dot{S}_{\beta) j} n^j + O \left( \frac{1}{\bar{r}^2} \right) + \bar{t}
\ O\left( \frac{1}{\bar{r}^3} \right) \,\, .
\end{equation}
It is clear that the terms of order $1/\bar{r}^2$ and
$\bar{t}/\bar{r}^3$ in $\bar{g}^{(1)}_{\bar{\alpha} \bar{\beta}}$
cannot contribute to the linearized Ricci tensor to order
$1/\bar{r}^3$. Similarly, it is clear that the terms of order
$1/\bar{r}^2$ and higher in the background scaled metric cannot
contribute to the linearized Ricci tensor to order $1/\bar{r}^3$, so,
to order $1/\bar{r}^3$, we see that $\bar{g}^{(1)}_{\bar{\alpha}
  \bar{\beta}}$ satisfies the linearized Einstein equation about the
Schwarzschild metric. It is therefore useful to expand $\bar{g}^{(1)}_{\bar{\alpha}
  \bar{\beta}}$ in tensor spherical harmonics.

We obtain one very useful piece of information by extracting the
$\ell=1$, magnetic parity part of the linearized Ricci tensor that is
even under time reversal. On account of the symmetries of the
background Schwarzschild metric, only the $\ell=1$, magnetic parity, even
under time reversal part of the metric perturbation can contribute.
Now, a general $\ell=1$, symmetric (but not necessarily trace free) tensor
field $Q_{\alpha \beta}(t,r,\theta,\phi)$ can be expanded in tensor
spherical harmonics as (see, e.g., \cite{thorne} or 
\cite{blanchet-damour} equations
(A16-18))
\begin{equation}\label{eq:harmonics}
\begin{split}
Q_{00} &= Q^A_i n^i \\
Q_{i0} &= Q^B_j n^j n_i + Q^C_i +  Q^M_k \epsilon_{i j}^{\ \ k} n^j \\
Q_{ij} &= Q^D_k n^k n_i n_j + Q^E_{(i}n_{j)} + Q^F_k \delta_{ij}n^k+ Q^N_k \epsilon^k_{\ l ( i}n_{j)}n^l,
\end{split}
\end{equation}
where the expansion coefficients
$Q^A_i,Q^B_i,Q^C_i,Q^D_i,Q^E_i,Q^F_i,Q^M_i,Q^N_i$ are functions of
$(t,r)$. 
The three-vector index on these coefficients
corresponds to the three different ``$m$-values'' for each $\ell=1$ harmonic.
Thus, we see that there are a grand total of eight types of $\ell=1$ tensor
spherical harmonics. The six harmonics associated with labeling indices
$A-F$ are of electric parity, whereas the two harmonics associated with
$M,N$ are of magnetic parity.

For the metric perturbation (\ref{eq:scaled-1}), the ``constant
tensors'' $h^{\textrm{\tiny tail}}_{\alpha \beta}$ and $A_{(\alpha ,
  \beta)}$ are purely electric parity and cannot contribute. It turns
out that $\frac{1}{\bar{r}}K_{\alpha \beta}(\theta,\phi)$ also does
not contribute to the $\ell=1$, magnetic parity part of the linearized
Ricci tensor that is even under time reversal: Since $K_{\alpha
  \beta}$ is independent of $\bar{t}$ the ``$M$'' part of $K_{\alpha
  \beta}$ is odd under time reversal, whereas the ``$N$'' part of
$\frac{1}{\bar{r}}K_{\alpha \beta}(\theta,\phi)$ is pure gauge.
Thus, the only term that contributes to order
$1/\bar{r}^3$ to the $\ell=1$, magnetic parity part of of the linearized
Ricci tensor that is even under time reversal is
$\frac{-4 \bar{t}}{\bar{r}^2} t_{(\alpha} \dot{S}_{\beta) j}
n^j$. Satisfaction of vacuum linearized Einstein equation at order
$1/\bar{r}^3$ requires that this term vanish. We thereby learn that
\begin{equation}
\frac{d S_{ij}}{dt} = 0 \,\, ,
\label{dSdt}
\end{equation}
i.e., to lowest order, the spin is parallelly propagated with respect
to the background metric along the worldline $\gamma$.

Having set the spin term to zero in eq.(\ref{eq:scaled-1}), 
we may now substitute the remaining
terms in eq. (\ref{eq:scaled-1}) into the linearized Einstein equation
and set the $1/\bar{r}^3$ part equal to zero. It is clear that we will
thereby obtain relations between $h^{\textrm{\tiny tail}}_{\alpha
  \beta}$, $A_{(\alpha , \beta)}$, and $K_{\alpha \beta}$. However,
these relations will not be of direct interest for obtaining
``equations of motion''---i.e., equations relating $A^i$ and its time
derivatives to known quantities---because the quantity of
interest $A_{i,0}$ always appears in combination with the quantity
$A_{0,i}$, which is unrelated to the motion. Therefore, we shall not explicitly compute the relations arising from the linearized Einstein equation here.

We now consider the information on $A^i$ that can be obtained from the
near zone second-order Einstein equation
\begin{equation}\label{eq:e2new}
G^{(1)}_{ab}[\bar{g}^{(2)}] = - G^{(2)}_{ab}[\bar{g}^{(1)}, \bar{g}^{(1)}],
\end{equation}
where, from eq.~(\ref{eq:scaled-metric2}), we see that
\begin{equation}\label{eq:scaled-2}
\begin{split}
\bar{g}^{(2)}_{\bar{\alpha} \bar{\beta}} &= B_{\alpha i \beta j} \bar{x}^i \bar{x}^j +
D_{\alpha \beta \gamma} \bar{x}^\gamma + M \mathcal{R}_{\alpha
  \beta}(\bar{x}^\mu) + H_{\alpha \beta}(\theta,\phi) +
\frac{\bar{t}}{\bar{r}} \dot{K}_{\alpha \beta}(\theta,\phi) \\ &+
O\left( \frac{1}{\bar{r}} \right) + \bar{t} O\left(
\frac{1}{\bar{r}^2} \right) + \bar{t}^2 O\left( \frac{1}{\bar{r}^3}
\right) \,\, ,
\end{split}
\end{equation}
where we have defined
\begin{align}
D_{\alpha \beta 0} & \equiv h^{\textrm{\tiny tail}}_{\alpha \beta 0} + 2 A_{(\alpha , \beta) 0} \\
D_{\alpha \beta i} & \equiv h^{\textrm{\tiny tail}}_{\alpha \beta i} + 2 A_{(\alpha , \beta) i} + 2 B_{\alpha i \beta j}A^j \,\, .
\end{align}
We wish to impose the second order Einstein equation to orders
$1/\bar{r}^2$ and $\bar{t}/\bar{r}^3$, which, as we shall see below,
are the lowest nontrivial orders in $1/\bar{r}$ as $\bar{r}
\rightarrow \infty$ that occur. First, we consider
$G^{(2)}_{ab}[\bar{g}^{(1)}, \bar{g}^{(1)}]$. The terms appearing in
this quantity can be organized into terms of the following general forms (i)
$\bar{g}^{(1)} \partial \partial \bar{g}^{(1)}$; (ii) $\partial
\bar{g}^{(1)} \partial \bar{g}^{(1)}$; (iii) $\Gamma \bar{g}^{(1)}
\partial \bar{g}^{(1)}$ where $\Gamma$ denotes a Christoffel symbol of
the background scaled metric; (iv) $\Gamma \Gamma
\bar{g}^{(1)}\bar{g}^{(1)}$; and (v) $(\partial
\Gamma)\bar{g}^{(1)}\bar{g}^{(1)}$.  From the form of $\bar{g}^{(1)}$
together with the fact that $\Gamma = O(1/\bar{r}^2)$ and
$\partial \Gamma = O(1/\bar{r}^3)$,
it is clear that none of these terms can contribute to
$G^{(2)}_{ab}[\bar{g}^{(1)},\bar{g}^{(1)}]$ to order $1/\bar{r}^2$ or
$\bar{t}/\bar{r}^3$. Therefore, we may treat $\bar{g}^{(2)}$ as satisfying
the homogeneous, vacuum linearized Einstein equation.

We now consider the linearized Ricci tensor of the perturbation
$\bar{g}^{(2)}$. By inspection of eq.~(\ref{eq:scaled-2}), it might
appear that terms that are $O(1)$ (from two partial derivatives acting
on the ``B'' term) and $O(1/\bar{r})$ (from various terms) will
arise. However, it is not difficult to show that the total
contribution to the $O(1)$ and $O(1/\bar{r})$ terms will vanish by
virtue of the fact that the metric $g_{ab}(\lambda=0)$ is a solution
to Einstein's equation and the term proportional to $\lambda$ in
eq.~(\ref{eq:metric2}) satisfies the far zone linearized Einstein
equation (which has already been imposed). It also is clear that there
is no contribution of $\bar{g}^{(2)}$ to the linearized Ricci tensor
that is of order $\bar{t}/\bar{r}^2$. Thus, the lowest nontrivial
orders that arise in the second order Einstein equation are indeed
$1/\bar{r}^2$ and $\bar{t}/\bar{r}^3$, as claimed above.

The computation of the linearized Ricci tensor to orders $1/\bar{r}^2$
and $\bar{t}/\bar{r}^3$ for the metric perturbation $\bar{g}^{(2)}$ is
quite complicated, so we will save considerable labor by focusing on
the relevant part of the linearized Einstein equation to these
orders. Our hope/expectation (which will be borne out by our
calculation) is to obtain an equation for ${A^i}_{,00}$. Since this
quantity is of $\ell=1$, electric parity type and is even under time
reversal, we shall focus on the $\ell=1$, electric parity, even under
time reversal part of the linearized Ricci tensor of $\bar{g}^{(2)}$
at orders $1/\bar{r}^2$ and $\bar{t}/\bar{r}^3$. From
eq.~(\ref{eq:harmonics}), we see that the $\ell=1$ electric parity 
part of the Ricci tensor
that is $O(1/\bar{r}^2)$ and even under time reversal can be written as
\begin{align}
R^{(1)}_{00}|_{\ell=1,+,\frac{1}{\bar{r}^2}} & = \frac{1}{\bar{r}^2} R^A_i n^i \\
R^{(1)}_{ij}|_{\ell=1,+,\frac{1}{\bar{r}^2}} & = \frac{1}{\bar{r}^2} \left( R^D_k n^k n_i n_j + R^E_{(i}n_{j)} + R^F_k n^k \delta_{ij} \right) \,\, ,
\end{align}
whereas the 
$\ell=1$ part of the Ricci tensor
that is $O(\bar{t}/\bar{r}^3)$ and even under time reversal can be written as
\begin{equation}
R^{(1)}_{i0}|_{\ell=1,+,\frac{\bar{t}}{\bar{r}^3}} = \frac{\bar{t}}{\bar{r}^3} \left( R^B_j n^j n_i +
R^C_i \right) \,\, .
\end{equation}
Here, in contrast to the usage of \eqref{eq:harmonics}, $R^A_i,R^B_i,R^C_i,R^D_i,R^E_i,R^F_i$ are ``constants'', i.e, they have
no dependence on $(\bar{t}, \bar{r})$.

We now consider the terms in $\bar{g}^{(2)}$ that can contribute to these
Ricci terms. The term $B_{\alpha i \beta j} \bar{x}^i \bar{x}^j$ has no
$\ell = 1$ part. Nevertheless, the $\ell = 2$ magnetic parity part of this
term can, in effect, combine with the $\ell=1$ magnetic parity ``spin term''
$\frac{1}{\bar{r}^2} t_{(\alpha} S_{\beta) j} n^j$ in the background scaled metric
to produce a contribution to the linearized Ricci tensor of the correct type.
This contribution will be proportional to
\begin{equation}\label{eq:spin-force-vector}
F_i \equiv S^{kl} R_{kl0i} \,\, .
\end{equation}

For the remaining terms in $\bar{g}^{(2)}$, the ``spin term''
$\frac{1}{\bar{r}^2} t_{(\alpha} S_{\beta) j} n^j$ in the background
scaled metric will not contribute to the relevant parts of the
linearized Ricci tensor, so we may treat the remaining terms in
$\bar{g}^{(2)}$ as though they were perturbations of
Schwarzschild. Thus, only the $\ell=1$, electric parity, even under
time reversal part of these terms can contribute.  The remaining contributors to $R^A_i,R^B_i,R^C_i,R^D_i,R^E_i$, and $R^F_i$ are
\begin{equation}
\begin{split}
D_{00k}\bar{x}^k & = \bar{r} D^A_in^i \\
D_{i00}\bar{t} & = \bar{t} D^C_i \\
D_{ijk}\bar{x}^k|_{\ell=1,+} & = \bar{r}( n_{(i}D^E_{j)} + \delta_{ij} n^k D^F_k) \,\, ,
\label{D}
\end{split}
\end{equation} and
\begin{align}
H_{00}|_{\ell=1,+} & = H^A_i n^i \\
\dot{K}_{i0}|_{\ell=1,+}  & = \dot{K}^B_j n^j n_i + \dot{K}^C_i \\
H_{ij}|_{\ell=1,+}  & = H^D_k n^k n_i n_j + H^E_{(i}n_{j)} + H^F_k n^k \delta_{ij},
\end{align}
where, in eq.~(\ref{D}), we have 
\begin{align}
D^A_i & = D_{00i}\\
D^C_i & = D_{i00}\\
D^E_i & = \frac{1}{5} \left( 3 D_{i \ k}^{\ k} - D^k_{\ k i} \right) \\
D^F_i & = \frac{1}{5} \left( - D_{i \ k}^{\ k} + 2 D^k_{\ k i} \right).
\end{align}(The curvature term $\mathcal{R}_{\alpha \beta}$ has not appeared in
the above equations because it has no $\ell=1$ part.)  The $D^A_i,D^C_i,D^E_i,D^F_i,H^A_i,H^D_i,H^E_i,H^F_i,\dot{K}^B_i,\dot{K}^C_i$ are also ``constants" in these expressions. A lengthy
calculation now yields

\setcounter{MaxMatrixCols}{15}

\begin{equation}\label{eq:matrix-of-doom}
\begin{pmatrix}
R^A_i \\ R^B_i \\ R^C_i \\ R^D_i \\ R^E_i \\ R^F_i \\
\end{pmatrix}
=
-\frac{1}{2}
\begin{pmatrix}
\frac{-16}{5} & -3M & 0   & -M  & -M  & -2 & 0  & 0 & 0  & -2 & 2  \\
0             & 0   & -6M & 0   & 0   & 0  & 0  & 0 & 0  & -3 & -3 \\
0             & 0   & 2M  & 0   & 0   & 0  & 0  & 0 & 0  & 1  & 1  \\
\frac{-6}{5}  & -6M & 0   & 0   & -6M & -3 & 3  & 0 & 3  & -6 & 0  \\
\frac{-16}{5} & 0   & 4M  & 0   & 4M  & 2  & -2 & 0 & -2 & 2  & -2 \\
2             & 0   & 3M  & -3M & 9M  & 1  & -3 & 0 & -3 & 2  & 0  \\
\end{pmatrix}
\begin{pmatrix}
F_i \\ D^A_i \\ D^C_i \\ D^E_i \\ D^F_i \\ H^A_i \\ H^D_i \\ H^E_i \\ H^F_i \\ \dot{K}^B_i \\ \dot{K}^C_i \\
\end{pmatrix}.
\end{equation}

Using the vacuum linearized Einstein equation $R^{(1)}_{ab}=0$, we
thus obtain 6 linear equations for our 11 unknowns. However, in order
to find ``universial'' behavior, we are interested in relations that
do not involve $H_{\alpha \beta}$ and $\dot{K}_{\alpha \beta}$. It can
be shown that there are two such relations\footnote{There will be
  three such relations in total, because the vanishing of the mass
  dipole moment for all time implies through $O(\lambda^2)$ in
  near-zone perturbation theory the vanishing of the value, time
  derivative, and second time
derivative of the mass dipole at time $t_0$.
  The third condition should follow from the first-order near-zone
  Einstein equation, which we did not fully use.  In fact, it should only be necessary
  to impose that the mass dipole have no second time derivative in order to
  define the motion.}, namely
\begin{equation}
-4 F_i - 3M D^A_i + 2 M D^C_i - 2 M D^E_i + 4 M D^F_i = 0 \,\, ,
\end{equation}
and
\begin{equation}
-F_i - M D^A_i + 2 M D^C_i = 0 \,\, .
\label{motion}
\end{equation}
The first equation involves $A^0$ and spatial derivatives of $A^i$,
and does not yield restrictions on the motion. However, the second
equation gives the desired equations of motion. Plugging in the definitions of the
quantities appearing in eq.~(\ref{motion}), we obtain
\begin{equation}
- S^{kl} R_{kl0i} - M ( h^{\textrm{\tiny tail}}_{00,i} + 2 R_{0 j
  0 i}A^j + 2 A_{0,0i} ) + 2 M ( h^{\textrm{\tiny tail}}_{i0,0} +
A_{i,00} + A_{0,i0} ) = 0,
\end{equation}
where we have taken advantage of the fact (noted above) that for $00$ and $0i$ components we have $h^{\textrm{\tiny tail}}_{\alpha \beta \gamma} = h^{\textrm{\tiny tail}}_{\alpha \beta , \gamma}$.  Using the equality of mixed partials $A_{0,i0}=A_{0,0i}$, we obtain
\begin{equation}
A_{i,00} = \frac{1}{2M} S^{kl} R_{kl0i} - R_{0 j 0 i}A^j - \left( h^{\textrm{\tiny tail}}_{i0,0} - \frac{1}{2} h^{\textrm{\tiny tail}}_{00,i} \right)\,\, .
\end{equation}
Thus, according to the interpretation provided in section
\ref{sec:dipole} above, the first order perturbative correction, $Z^i(t)$,
to the geodesic $\gamma$ of the background spacetime satisfies
\begin{equation}\label{eq:EOM}
\frac{d^2Z^i}{dt^2} = \frac{1}{2M} S^{kl} {R_{kl0}}^i - {R_{0j0}}^iZ^j
- \left( {h^{\textrm{\tiny tail}}}^i{}_{0,0} - \frac{1}{2}
h^{\textrm{\tiny tail}}_{\ \ \ 00}{}^{,i} \right) \,\, .
\end{equation}
In addition, we have previously found that $M$ and $S_{ij}$ are constant
along $\gamma$. Taking account of the fact that this equation is written
in Fermi normal coordinates of $\gamma$ and that $Z^0 = 0$,
we can rewrite this equation in a more manifestly covariant looking form
as
\begin{equation}\label{eq:EOMcov}
u^c\nabla_c(u^b\nabla_b Z^a) = \frac{1}{2M} {R_{bcd}}^a S^{bc} u^d 
- {R_{bcd}}^au^bZ^cu^d
- (g^{ab} + u^a u^b)(\nabla_d
h_{bc}^{\tiny \textrm{tail}}- \frac{1}{2} \nabla_b h_{cd}^{\tiny
\textrm{tail}})u^c u^d \,\, .
\end{equation}
where $u^a$ is the tangent to $\gamma$ and $u^c \nabla_c S_{ab} = 0$.
However, it should be emphasized that this equation describes the
perturbed motion only when the metric perturbation is in the Lorenz gauge
(see Appendix A).

The first term in eq.~(\ref{eq:EOM}) (or, equivalently, in
eq.~(\ref{eq:EOMcov})) is the ``spin force'' first obtained by Papapetrou
\cite{papapetrou}. Contributions from higher multipole moments do not
appear in our equation because they scale to zero faster than the spin
dipole moment, and thus would arise at higher order in $\lambda$ in
our perturbation scheme.  The second term corresponds to the right
side of the geodesic deviation equation, and appears because the
perturbed worldline is not (except at special points) coincident with
the background worldline\footnote{Consider a one-parameter-family wherein the initial position for a body is ``moved over'' smoothly with $\lambda$.  In the limit $M \rightarrow 0$, the body then moves on a family of geodesics of the background metric parameterized by $\lambda$, and the perturbative description of motion should indeed be the geodesic deviation equation.}. The final term is the ``gravitational
self-force'', which is seen to take the form of a (regularized)
gravitational force from the particle's own field. Our derivation has
thus provided a rigorous justification of the regularization schemes
that have been proposed elsewhere.

Finally, we note that, although our analysis has many points of
contact with previous analyses using matched asymptotic expansions,
there are a number of significant differences. We have already noted
in section \ref{sec:geodesic} that our derivation of geodesic motion
at zeroth order in $\lambda$ appears to differ from some other
derivations \cite{d'eath,mino-sasaki-tanaka,poisson}, which do not
appear to impose the requirement that $M \neq 0$. We also have
already noted that in other approaches to self-force
\cite{mino-sasaki-tanaka,poisson}, what corresponds to our scaled
metric at $\lambda = 0$ is {\it assumed} to be of Schwarzschild form.
In these other approaches, first order perturbations in the near zone
expansion are treated as time independent, and are required to be regular on the Schwarzschild horizon. By contrast we make no assumptions about the time-dependence of
the perturbations of the scaled metric beyond those that follow from
our fundamental assumptions (i)-(iii) of section \ref{sec:example}.
Thus, our first order perturbations are allowed to have linear
dependence on $\bar{t}$, and our second order perturbations can depend
quadratically on $\bar{t}$. We also make no assumptions about the
spacetime at $\bar{r} < \bar{R}$ and therefore impose no boundary
conditions at small $\bar{r}$.  Finally, there is a significant
difference in the manner in which the gauge conditions used to define the 
motion are imposed.  In
\cite{mino-sasaki-tanaka,poisson}, the entire $\ell = 1$ electric parity
part of what corresponds to our second order near zone perturbation is
set to zero without proper justification\footnote{Note that the part of the $\ell=1$ electric parity perturbation that is relevant for obtaining equations of motion
in \cite{mino-sasaki-tanaka,poisson} is of ``acceleration type'' (with
linear growth in $\bar{r}$) and does not have an obvious interpretation
in terms of a shift in the center of mass.}.
By contrast, our ``no mass dipole''
condition
applies to the \textit{background} near-zone metric and has been justified
as providing ``center of mass'' coordinates.

\section{Beyond Perturbation Theory}\label{sec:beyond}

As already mentioned near the beginning of section \ref{sec:dipole},
the quantity $Z^i$ in eq.~\eqref{eq:EOM} is a ``deviation vector''
defined on the background geodesic $\gamma$ that describes the first
order in $\lambda$ perturbation to the motion. For any one parameter
family of spacetimes $g_{ab}(\lambda)$ satisfying the assumptions
stated in section \ref{sec:example}, eq.~\eqref{eq:EOM} is therefore
guaranteed to give a good approximation to the deviation from the
background geodesic motion $\gamma$ as $\lambda \rightarrow 0$. In
other words, if $\gamma$ is described by $x^i(t) = 0$, then the new
worldline obtained defined by $x^i(t) = \lambda Z^i(t)$ is the correct
description of motion to first order in $\lambda$ (when the metric
perturbation is in Lorenz gauge) and is therefore guaranteed to be
accurate at small $\lambda$. However, this guarantee is of the form
that if one wants to describe the motion accurately up to time $t$,
then it always will be possible to choose $\lambda$ sufficiently small
that $x^i(t) = \lambda Z^i(t)$ is a good approximation up to time
$t$. The guarantee is {\it not} of the form that if $\lambda$ is
chosen to be sufficiently small, then $x^i(t) = \lambda Z^i(t)$ will
accurately describe the motion for all time.  Indeed, for any fixed
$\lambda>0$, it is to be expected that $Z^i(t)$ will grow large at
sufficiently late times, and it is clear that the approximate
description of motion $x^i(t) = \lambda Z^i(t)$ cannot be expected to
be good when $Z^i(t)$ is large, since by the time the motion has
deviated significantly from the original background geodesic $\gamma$,
the motion clearly cannot be accurately described in the framework of
being a ``small correction'' to $\gamma$. However, the main intended
application of the first order corrected equations of motion is to
compute motion in cases, such as inspiral, where the deviations from
the original geodesic motion become large at late times. It is
therefore clear that eq.~\eqref{eq:EOM}, as it stands, is useless for
computing long term effects, such as inspiral.

One possible response to the above difficulty would be to go to higher
order in perturbation theory. However, it seems clear that this will
not help. Although the equations of motion obtained from $n$th order
perturbation theory will be more accurate than the first order
equations, they will not have a domain of validity that is
significantly larger than the first order equations. The perturbative
description at any finite order will continue to treat the motion as a
``small deviation'' from $\gamma$, and cannot be expected to describe
motion accurately when the deviations are, in fact, large. In essence,
by the time that the deviation from $\gamma$ has become sufficiently
large to invalidate first order perturbation theory---so that, e.g.,
the second order corrections are comparable in magnitude to the first
order corrections---then one would expect that the $(n+1)$th order
corrections will also be comparable to the $n$th order corrections, so
$n$th order perturbation theory will not be accurate either. Only by
going to all orders in perturbation theory can one expect to get an
accurate, global in time, description of motion via perturbation
theory. Of course, if one goes to all orders in perturbation theory,
then there is little point in having done perturbation theory at all.

Nevertheless, for a sufficiently small body of suffciently small mass,
it seems clear that the corrections to geodesic motion should be {\it
  locally} small and should be locally described by
eq.~\eqref{eq:EOM}.  By the time these small corrections have built up
and the body has deviated significantly from the original geodesic
approximating its motion, it should then be close to a \textit{new}
geodesic, perturbing off of which should give a better approximation
to the motion for that portion of time. One could then attempt to
``patch together'' such solutions to construct a world-line that
accurately describes the motion of the particle for a longer time. In
the limit of many such patches with small times between them, one
expects the resulting worldline to be described by a single
``self-consistent''
differential equation, which should then well-approximate the motion
as long as it remains \textit{locally} close to geodesic motion.

A simple, familiar example will help illustrate all of the above
points. Consider the cooling of a ``black body''. To choose a definite
problem that can be put in a framework similar to that considered in
this paper, let us consider a body (such as a lump of hot coal) that
is put in a box with perfect reflecting walls, but a hole of area
$A$ is cut in the this wall. We are interested in determining how the
energy, $E$, of the body changes with time. At finite $A$, this is a
very difficult problem, since the body will not remain in exact
thermal equilibrium as it radiates energy through the hole. However,
let us consider a one-parameter family of cavities where $A(\lambda)$
smoothly goes to zero as $\lambda \rightarrow 0$. When $\lambda=0$, we
find that the energy, $E_0 \equiv E(\lambda=0)$, does not change with
time, and the body will remain in thermal equilibrium at temperature
$T_0$ for all time. When we do first order perturbation theory in
$\lambda$, we will find that the first order in $\lambda$ correction,
$E^{(1)}$, to the energy satisfies\footnote{Of course, when $A$
  becomes small compared to the typical wavelengths of the radiation
  (as it must as we let $A \rightarrow 0$), we enter a physical optics
  regime where our formulas are no longer valid. We ignore such
  effects here, just as in our above analysis of the motion of bodies
  in general relativity we ignored quantum gravity effects even though
  they should be important when the size of the body is smaller than
  the Planck scale.}
\begin{equation}
\frac{dE^{(1)}}{dt} = - \sigma A^{(1)} T^4_0
\label{dE}
\end{equation}
where $\sigma$ is the Stefan-Boltzmann constant and $A^{(1)} \equiv
dA/d\lambda |_{\lambda=0}$. Note that only the zeroth order
temperature, $T_0$, enters the right side of this equation because the
quantity $A^{(1)}$ is already first order in $\lambda$, so the effect
of any changes in temperature would appear only to higher order
in $\lambda$. Since $T_0$ is a constant, it is easy to integrate
eq.~(\ref{dE}) to obtain,
\begin{equation}
E^{(1)}(t) = - \sigma A^{(1)} T^4_0 t
\label{E1t}
\end{equation}
Thus, first order perturbation theory approximates the behavior of
$E(\lambda, t)$ as 
\begin{equation}
E(\lambda, t) = E_0 - \lambda \sigma A^{(1)} T^4_0 t
\label{Et}
\end{equation}
Although this is a good approximation at early times, it is a horrible
approximation at late times, as it predicts that the energy will go
negative. If one went to second order in perturbation theory, one
would obtain corrections to eq.~(\ref{dE}) that would take into
account the first order energy loss as well as various non-equilibrium
effects. However, one would still be perturbing off of the
non-radiating background, and the late time predictions using second
(or any finite higher order) perturbation theory would still be very poor.

However, there is an obvious major improvement that can be obtained by
noting that if $A$ is sufficiently small, then the body should remain
nearly in thermal equilibrium as it loses energy. Therefore, although
perturbation theory off of the zeroth order solution may give poor
results at late times, first order perturbation theory off of {\it
  some} thermal equilibrium solution should give locally accurate
results at all times. This suggests that if $A$ is sufficiently small,
the cooling of the body should be described by
\begin{equation}
\frac{dE}{dt} = - \sigma A T^4(t) \,\, .
\label{dEnew}
\end{equation}
When supplemented with the formula, $E=E(T)$, that relates energy to
temperature when the body is in thermal equilibrium, this equation
should provide an excellent description of the cooling of the body
that is valid at all times. In effect, eq.~(\ref{dEnew}) takes into
account the higher order perturbative effects (to all orders in
$\lambda$) associated with the cooling of the body, but it neglects
various perturbative effects associated with the body failing to
remain in thermal equilibrium as it cools. Equation (\ref{dEnew}) is
{\it not} an exact equation (since it does not take various
non-equilibrium effects into account) and it is {\it not} an equation
that arises directly from perturbation theory. Rather, it is an
equation that corresponds to applying first order perturbation theory
to a background that itself undergoes changes resulting from the
perturbation. We will refer to such an equation as a ``self-consistent
perturbative equation''. Such equations are commonly written down for
systems that can be described {\it locally in time} by a small
deviation from a simple solution.

How does one find a ``self-consistent perturbative equation'' for a
given system for which one has derived first order perturbative
equations? We do not believe that there is any general method for
deriving a self-consistent perturbative equation. However, the
following appear to be appropriate criteria to impose on a
self-consistent perturbative equation: (1) It should have a well posed
initial value formulation. (2) It should have the same number of
degrees of freedom as the first order perturbative system, so that a
correspondence can be made between initial data for the
self-consistent perturbative equation and the first order perturbative
system. (3) For corresponding initial data, the solutions to the
self-consistent perturbative equation should be close to the
corresponding solutions of the first order perturbative system over
the time interval for which the first order perturbative description
should be accurate. We do not know of any reason why, for any given
system, there need exist a self-consistent perturbative equation
satisfying these criteria. In cases where a self-consistent
perturbative equation satisfying these criteria does exist, we would
not expect it to be unique. For example, we could modify eq.~(\ref{dEnew})
by adding suitable terms proportional to $A^2$ to the right side of this
equation.

The first order perturbative equations for the motion of a small body 
are that the first order metric
perturbation satisfies
\begin{equation}
\nabla^c \nabla_c \tilde{h}_{ab} - 2 R^c{}_{ab}{}^d \tilde{h}_{cd} =
- 16 \pi M u_a u_b \frac{\delta^{(3)}(x^i)}{\sqrt{-g}} \frac{d\tau}{dt}\,\,,
\end{equation}
where $x^i = 0$ corresponds to a geodesic, $\gamma$ of the background
spacetime, and $u^a$ is the tangent to $\gamma$. If we consider the
retarded solution to this equation (which automatically satisfies the
Lorenz gauge condition), we have proven rigorously in this paper that
the first order in $\lambda$ deviation of the motion from $\gamma$
satisfies
\begin{equation}
u^c\nabla_c(u^b\nabla_b Z^a) =  
- {R_{bcd}}^au^bZ^cu^d
- (g^{ab} + u^a u^b)(\nabla_d
h_{bc}^{\tiny \textrm{tail}}- \frac{1}{2} \nabla_b h_{cd}^{\tiny
\textrm{tail}})u^c u^d \,\, ,
\end{equation}
with
\begin{equation}
h_{ab}^{\tiny \textrm{tail}}(x) = M
\int_{-\infty}^{\tau_{\textrm{ret}}^-}\left(G^+_{a b a'
b'}-\frac{1}{2}g_{ab}G^{+ \ c}_{\ c \ a ' b
'}\right) (x,z(\tau')) u^{a '}u^{b '} d\tau ' \,\, ,
\end{equation}
where, for simplicity, we have dropped the spin term. The MiSaTaQuWa 
equations
\begin{equation}
\nabla^c \nabla_c \tilde{h}_{ab} - 2 R^c{}_{ab}{}^d \tilde{h}_{cd} =
- 16 \pi M u_a(t) u_b(t) \frac{\delta^{(3)}(x^i - z^i(t))}{\sqrt{-g}} \frac{d\tau}{dt},
\label{misa2}
\end{equation}
\begin{equation}
u^b \nabla_b u^a = - (g^{ab} + u^a u^b)(\nabla_d
h_{bc}^{\tiny \textrm{tail}}- \frac{1}{2} \nabla_b h_{cd}^{\tiny
\textrm{tail}})u^c u^d \,\, ,
\label{motion2}
\end{equation}
\begin{equation}
h_{ab}^{\tiny \textrm{tail}}(x) = M
\int_{-\infty}^{\tau_{\textrm{ret}}^-}\left(G^+_{a b a'
b'}-\frac{1}{2}g_{ab}G^{+ \ c}_{\ c \ a ' b
'}\right)\left(x,z(\tau ')\right)u^{a '}u^{b '} d\tau ' \,\, ,
\end{equation}
(where one chooses the retarded solution to eq.~(\ref{misa2}))
are an excellent candidate for self-consistent perturbative equations
corresponding to the above first order perturbative
system\footnote{The Riemann tensor term does not appear on the right
  side of eq.~(\ref{motion2}), since in the self-consistent
  perturbative equation, the deviation from the self-consistent
  worldline should vanish.}.  Here, $u^a(\tau)$ (normalized in the background metric) refers to the self-consistent motion $z(\tau)$, rather than to a background geodesic as before.
Although a proper mathematical analysis of this integro-differential
system has not been carried out, it appears plausible that our above
criteria (1)-(3) will be satisfied by the MiSaTaQuWa equations. If so,
they should provide a good, global in time,
description of motion for problems like extreme mass ratio inspiral.

\begin{acknowledgments}
We wish to thank Abraham Harte and Eric Poisson for helpful discussions.  This research was supported in part by NSF grant PHY04-56619 to the University of Chicago and a National Science Foundation Graduate Research Fellowship to SG.
\end{acknowledgments}

\appendix
\section{Self-force in an Arbitrary Allowed Gauge}

As discussed in section \ref{sec:dipole}, the description of motion
will change under first-order changes of gauge. Indeed, in that section, we
noted that under a smooth gauge transformation, the description of
motion changes by eq.~(\ref{smoothgauge}). However, as previously
stated near the end of section \ref{sec:example} above (see equation \eqref{eq:mushi-mushi}), the allowed coordinate freedom
includes transformations that are not smooth at $r=0$. Since such
gauges may arise in practice\footnote{For example, the Regge-Wheeler gauge (used for perturbations of the Schwarzschild metric) is not smoothly related to the Lorenz gauge \cite{barack-ori}.  However, it is possible that the gauge vector is bounded \cite{barack-ori}, in which case perturbations in the Regge-Wheeler gauge would satisfy our assumptions (see equation \eqref{xiform}), and equations of motion could be defined.  On the other hand, point particle perturbations expressed in radiation gauges (used for perturbations of the Kerr metric) contain a $\log$ singularity along a string \cite{barack-ori}, and therefore do not satisfy our assumptions.}, we provide here the expression for the
first order perturbative equation of motion in an arbitrary gauge allowed by our assumptions.  We also present the corresponding self-consistent perturbative equations of motion.

As previously noted in section \ref{sec:calculation} (see the remark below
eq.~(\ref{jH})), the equations of motion to first order in $\lambda$ depend
only upon the first order gauge transformation $\xi^a$.
As we have seen, the mass dipole moment appears at second-order in
(far zone) perturbation theory, so we must consider the
effects of first-order gauge transformations on second-order
perturbations.  This is given by $g^{(2)} \rightarrow g^{(2)} + \delta
g^{(2)}$, with
\begin{equation}
\delta g^{(2)}_{ab} = (\mathcal{L}_\xi g^{(1)})_{ab} + \frac{1}{2}
(\mathcal{L}^2_\xi g^{(0)})_{ab} \,\, ,
\end{equation}
where $\mathcal{L}$ denotes the Lie derivative.  Equivalently, we have
\begin{equation}
\delta g^{(2)}_{ab} = \xi^c \nabla_c g^{(1)}_{ab} + 2
\nabla_{(a} \xi^c g^{(1)}_{b)c} + \xi^c \nabla_c \nabla_{(a} \xi_{b)} + \nabla_{(a} \xi^c \nabla_{b)} \xi_c  + \nabla_c \xi_{(a} \nabla_{b)} \xi^c \,\, ,
\label{dg2}
\end{equation}
where $\nabla_a$ is the derivative operator associated with the
background metric $g_{ab}(\lambda=0)$. In order to satisfy the
criteria on allowed gauge transformations (see equation \eqref{eq:mushi-mushi}), the components
of $\xi^a$ must be of the form
\begin{equation}
\xi^\mu = F^\mu(t,\theta, \phi) + O(r) \,\, ,
\label{xiform}
\end{equation}
i.e., $\xi^a$ cannot ``blow up'' at $r=0$ but it can be singular in the sense
that its components can have direction-dependent limits. [Note added: Please see arXiv:1104.5205.]

The mass dipole moment, $P^i$, is 
one-half of the coefficient of the
$\ell=1$ part of the leading order, $1/r^2$, part of the second order
metric perturbation, $g^{(2)}_{00}$.
Therefore, $P^i$ may be extracted from the formula,
\begin{equation}\label{eq:app-dip}
P^i = \frac{3}{8 \pi} \lim_{R \rightarrow 0}\int_{r=R} g^{(2)}_{00} n^i dS \,\, ,
\end{equation}
where dS is the area element on the sphere of radius $R$.
Under the gauge transformation generated by $\xi^a$, we have
\begin{equation}
\delta g^{(2)}_{00} = \xi^c \nabla_c g^{(1)}_{00} + 2
\nabla_{0} \xi^c g^{(1)}_{0c} + \xi^c \nabla_c \nabla_{0} \xi_{0} + \nabla_{0} \xi^c \nabla_{0} \xi_c  + \nabla_c \xi_{0} \nabla_{0} \xi^c \,\, .
\label{dg200}
\end{equation}
As previously noted, for 
an arbitrary first-order perturbation satisfying our
assumptions, we have
\begin{equation}
g^{(1)}_{00} = \frac{2M}{r} + O(1),
\label{g100}
\end{equation}
where $M$ is the mass of the body. From eqs.~(\ref{xiform}), (\ref{dg200}) and
(\ref{g100}), we see that the change in $g^{(2)}_{00}$
induced by
our gauge transformation is 
\begin{equation}\label{eq:delta-g2-0}
\delta g^{(2)}_{00} = -\frac{2M}{r^2} \xi^i n_i + O\left( \frac{1}{r} \right).
\end{equation}\addtocounter{equation}{1}
Therefore, by eq.~(\ref{eq:app-dip}), the induced change in the mass dipole
moment is
\begin{equation}
\delta P^i = \frac{-3M}{4 \pi} \lim_{r \rightarrow 0}\int \xi^j n_j n^i d\Omega \,\, ,
\label{dP}
\end{equation}
where $d\Omega$ is the area element on the unit sphere.

Equation (\ref{dP}) gives the change in the mass dipole moment induced by
the possibly non-smooth gauge transformation generated by $\xi^a$. The
corresponding change in the first order perturbative equation of
motion is determined by the change in the
\textit{smooth} vector field $A^a$ required to eliminate the mass dipole.
Writing $A^a \rightarrow A^a + \delta A^a$, this change is given by
\begin{equation}
\delta A^i = \delta P^i/M 
\end{equation}
(see eq.\eqref{AP}). Thus, the
change $Z^i \rightarrow \hat{Z}^i = Z^i + \delta Z^i$ 
induced in the deviation vector
describing the perturbed worldline is
\begin{equation}\label{eq:gauge-law}
\delta Z^i = \frac{-3}{4 \pi} \lim_{r \rightarrow 0}\int \xi^j n_j  n^i d\Omega.
\end{equation}
In the case where our original gauge
was the Lorenz gauge, it follows immediately 
from eq.~(\ref{eq:EOM}) that the
new equation of motion for $\hat{Z}^i$ is
\begin{equation}\label{eq:EOM-app2}
\frac{d^2 \hat{Z}^i}{dt^2} = - {R_{0j0}}^iZ^j
- \left( {h^{\textrm{\tiny tail}}}^i{}_{0,0} - \frac{1}{2}
h^{\textrm{\tiny tail}}_{\ \ \ 00}{}^{,i} \right) + \ddot{\delta Z}^i \,\, ,
\end{equation}
where $\delta Z^i$ is given by eq.~(\ref{eq:gauge-law}), and where, for simplicity,
we have dropped the spin term. We may rewrite eq.~(\ref{eq:EOM-app2}) as
\begin{equation}\label{eq:EOM-app6}
\frac{d^2 \hat{Z}^i}{dt^2} = - {R_{0j0}}^i\hat{Z}^j
- \left( {h^{\textrm{\tiny tail}}}^i{}_{0,0} - \frac{1}{2}
h^{\textrm{\tiny tail}}_{\ \ \ 00}{}^{,i} \right) + \ddot{\delta Z}^i 
+ {R_{0j0}}^i {\delta Z}^j \,\, .
\end{equation}
Note that although eq.~(\ref{eq:EOM-app6}) provides us with the desired
equation of motion in an arbitrary allowed gauge, the terms involving
components of $h^{\textrm{\tiny tail}}$ must still be computed in the Lorenz
gauge.

Now suppose one wishes to pass to a self-consistent perturbative
equation associated with the new choice of gauge. It is not obvious how one might wish to modify the evolution equations for the
metric perturbations in the new gauge. (One possibility would be to
simply use eq.~(\ref{misa2}) and then modify the result by the
addition of $2\nabla_{(a} \xi_{b)}$ but it might be preferable to find
a new equation based on a suitable ``relaxed'' version of the
linearized Einstein equation for the new gauge.) However, it appears
that a natural choice of self-consistent perturbative equation
associated to eq.~(\ref{eq:EOM-app6}) would be
\begin{equation}
u^b \nabla_b u^a = - (g^{ab} + u^a u^b)(\nabla_d
h_{bc}^{\tiny \textrm{tail}}- \frac{1}{2} \nabla_b h_{cd}^{\tiny
\textrm{tail}})u^c u^d + \ddot{\delta Z}^a 
+ {R_{cbd}}^a u^c u^d {\delta Z}^b  \,\, .
\end{equation}
In the case where $\xi^a$ is smooth (so that, by eq.~(\ref{eq:gauge-law}),
we have $\delta Z^i = - \xi^i$) this agrees with the proposal of
Barack and Ori \cite{barack-ori}.

\end{document}